\documentclass[aps,prapplied,reprint,groupedaddress,longbibliography]{revtex4-2} 
\usepackage{graphicx}
\usepackage{dcolumn}
\usepackage{tikz}

\usepackage{bm}
\usepackage{graphicx}
\usepackage[caption=false]{subfig}

\usepackage[colorlinks=true, citecolor=blue, linkcolor=blue, urlcolor=blue]{hyperref}
\usepackage{pinlabel} 
\usepackage{amsmath}
\usepackage{dcolumn}
\usepackage{xcolor}

\usepackage{longtable}
\usepackage{amssymb}
\usepackage{natbib}
\usepackage{mathrsfs}
\usepackage[T1]{fontenc}
\definecolor{pakistangreen}{rgb}{0.0, 0.4, 0.0}
\definecolor{phthalogreen}{rgb}{0.07, 0.21, 0.14}
\newcommand{\appropto}{\mathrel{\vcenter{\offinterlineskip\halign{\hfil$##$\cr\propto\cr\noalign{\kern2pt}\sim\cr\noalign{\kern-2pt}}}}}

\newcommand{\beq}{\begin{equation}}
\newcommand{\eeq}{\end{equation}}
\newcommand{\beqr}{\begin{eqnarray}}
\newcommand{\eeqr}{\end{eqnarray}}

\newcommand{\wi}{\omega_c^{\rm I}}
\newcommand{\wia}{\omega_c^{\rm GI}}
\newcommand{\wib}{\omega_c^{\rm AI}}

\newcommand{\wE}{\omega_c^{\rm E}}

\newcommand{\RN}[1]{%
  \textup{\uppercase\expandafter{\romannumeral#1}}%
}
\newcommand{\Hdc}{H^{(c)}_{d}}
\newcommand{\Hdnc}{H^{(nc)}_{d}}

\usepackage{color}

\begin{document}


\title{An error-protected cross-resonance switch in weakly-tuneable architectures}
\author{Xuexin Xu}
\affiliation{Institute for Quantum Information,
RWTH Aachen University, D-52056 Aachen, Germany}
\affiliation{Peter Gr\"unberg Institute, Forschungszentrum J\"ulich, J\"ulich 52428, Germany}
\author{M. Ansari}
\affiliation{Institute for Quantum Information,
RWTH Aachen University, D-52056 Aachen, Germany}
\affiliation{Peter Gr\"unberg Institute, Forschungszentrum J\"ulich, J\"ulich 52428, Germany}
\begin{abstract}
In two-qubit gates activated by microwave pulses, by turning pulse on or off, the state of qubits are swapped between entangled or idle modes. In either mode, the presence of stray couplings makes qubits accumulate coherent phase error. However, the error rates in the two modes differ because qubits carry different stray coupling strengths in each mode; therefore, eliminating stray coupling from one mode does not remove it from the other. We propose to combine such a gate with a tunable coupler and show that both idle and entangled qubits can become free from stray couplings. This significantly increases the operational switch fidelity in quantum algorithms.  We further propose a weakly-tunable qubit as an optimum coupler to bring the two modes parametrically near each other. This remarkably enhances the tuning process by reducing its leakage. 

\end{abstract}


\maketitle


\section{Introduction} 
\vspace{-0.1in}
Over a few decades, quantum computing has evolved from a concept~\cite{feynman1982simulating} to experiments on noisy intermediate-scale quantum processors ~\cite{preskill2018quantum,bharti2021noisy,linghu2022quantum}. In the processors, entangled qubits together search for the answer state to a computational problem and this outperforms classical computational time ~\cite{arute2019quantum,foxen2020demonstrating,wu2021strong}. In gate-based quantum processors, quantum maps are decomposed into a sequence of one and two-qubit gates. Engineering these gates has targeted fast and less erroneous quantum state change. 

Scaling of current superconducting processors towards thousands of qubits is in principle possible; however, the key obstacle for current state-of-the-art technology is the large gate error rates~\cite{harper2019fault-tolerant,chen2021exponential,noiri2022fast}. The two-qubit gate error rate has been significantly reduced over the past few years and is currently about $0.4$ percent, one error per 250 operations~\cite{sung2021realization, jurcevic2021demonstration, huang2022quantum}. However, such error rates are far from useful quantum advantage and error correction.  There is a clear knowledge gap in the literature on enabling lower qubit error rates. While careful control of material noise sources, such as trap states~\cite{ansari2011noise, tian2021a-josephson} and nonequilibrium quasiparticle tunnelling~\cite{serniak2018hot-nonequilibrium, ansari2015rate, bal2015dynamics}, improves qubit coherence time and leads to a lower gate error rate, however, gates should be protected against non-material sources of errors, such as leakage to or from non-computational levels, and crosstalk due to stray couplings between interacting qubits.   

There is still a significant unwanted stray coupling between two idle qubits, which leads to an always-on and accumulating correlated coherent error for non-gated qubits when running quantum algorithms~\cite{krinner2020benchmarking}. A common solution to eliminate stray crosstalk is introducing a tunable coupling element, investigated by numerous groups~\cite{collodo2020implementation,xu2020high-fidelity, moskalenko2022high}. The same tunable coupler between two qubits may also be used to find an operating point for introducing strong two qubits interaction in a controllable way, which relates the tunable coupler to the concept of ON/OFF switch in CZ gates~\cite{sung2021realization}.  

Next to the concept of gates that require a tunable coupler between two qubits, there is a second class of gates more suitable for circuits with fixed-frequency qubits and couplers. In the latter class, a commonly used gate is the so-called cross-resonance (CR) gate, which is realized by applying a microwave pulse on a qubit and driving it with the frequency of another qubit. The CR gate supplies weak yet fully controllable $ ZX$ interaction between the two qubits for as long as the pulse is applied. Single qubit gates can follow it to introduce a CNOT operation on the initial two-qubit state.   In state-of-the-art circuits, the qubit-qubit stray $ZZ$ coupling is 50-100 times weaker than the qubit-qubit engineered coupling strength; however, this reduces CR gate fidelity by~1$\%$~\cite{krinner2020demonstration,kandala2021demonstration,zhao2022quantum}. Recently remarkable advances for better CR gates have been proposed by zeroing stray coupling in a hybrid circuit ~\cite{ku2020suppression,zhao2020high-contrast,jin2021implementing}, and between two transmons~\cite{sung2021realization,ni2021scalable,sete2021parametric-resonance}. 

A problem with the state-of-the-art CR gate is that switching between idle and entanglement is achieved by turning on and off the microwave pulse. This is a problem because the stray coupling strength changes in the presence or absence of a microwave pulse. For example, one can use the protocol called `dynamic $ZZ$-freedom' --- see section \RN{5} of Ref. \cite{xu2021zz-freedom, ansari2022method} ---  to eliminate total stray coupling in the presence of CR pulse; however, once the pulse is switched off, the idle qubits start to feel some residual stray coupling, making them accumulate coherent phase error. Similar is the other way around: once stray couplings are eliminated in the absence of microwave pulse using `static $ZZ$-freedom' protocol --- in section \RN{3} of Ref. \cite{xu2021zz-freedom, ansari2021circuit} --- idle qubits perform better, however activating CR gate introduces some risidual $ZZ$ coupling to lower CR gate fidelity.

Here we introduce a parasitic-free cross-resonance on/off switch, namely parasitic-free (PF) gate, which eliminates $ZZ$ stray coupling from both idle and entangled operating modes. We study the characteristics of this gate by deep modelling it. We show that the required parasitic-freedom feature occurs by tuning the coupling strength in the presence or absence of a microwave pulse.  We show several operating points for switching the CR gate on and off without leaving any parasitic $ZZ$ error on the states.  Among all operating points, we narrow our search for the most optimum on/off points to be the closest to one another, and for such points, we work out the most optimum tuning pulse for minimal leakage.

\begin{figure}[t]
	\centering			
\hspace{0.35in}\includegraphics[width=0.45\textwidth]{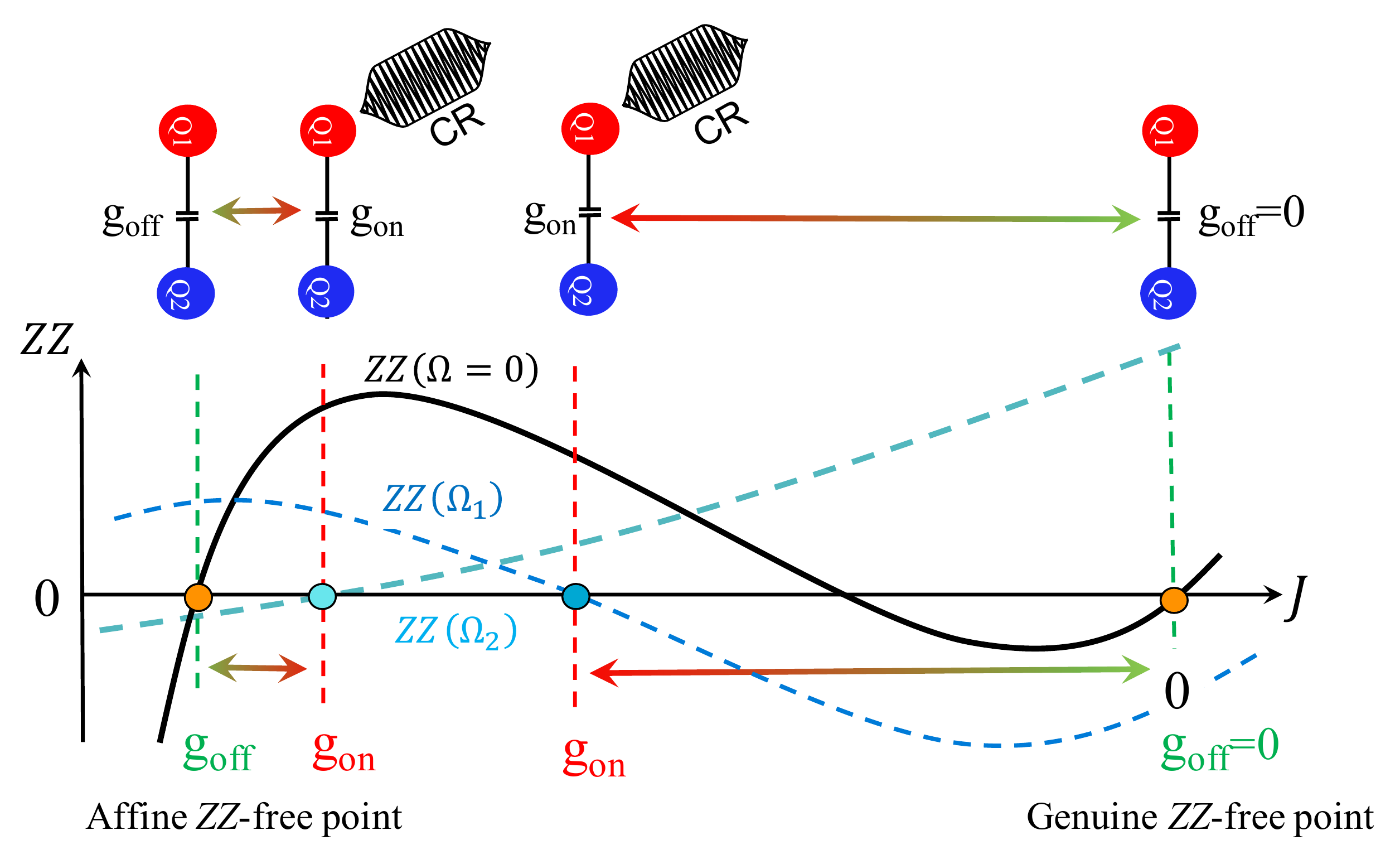}\\
		\vspace{-0.15in}
	\caption{Schematic PF gate. At $g_{\rm{off}}$ ($g_{\rm{on}}$) points, the qubits are at idle (entangled) mode with net zero parasitic $ZZ(\Omega)$ interaction. The static $ZZ$ interaction, shown as $ZZ(\Omega=0)$, has zeros at genuine and affine points.}\label{fig:diagram}
	\vspace{-0.1in}
\end{figure}

Figure~\ref{fig:diagram} shows the summary of our proposal. Different from the recently implemented CR gate~\cite{cai2021impact}, our PF gate can switch between idle (I) and entangled (E) modes without collecting any residual $ZZ$ in the modes. Residual interaction in the presence of CR drive is denoted by $ZZ(\Omega)$ with $\Omega$ being the microwave pulse amplitude and $ZZ(0)$ indicating static stray coupling in the absence of microwave. Switching mainly occurs by enabling tunability in coupling strength $g$ between qubits. At idle mode, as long as the coupler is parked at  $g_{\rm{off}}$, the qubits do not pick up the erroneous phase. Approaching such a parasitic-free switch is an essential step toward error-free quantum computation.

\section{Principles} \vspace{-0.1in}\label{sec:principle}
The parasitic-free (PF) gate is a variant of CR gate with zero $ZZ$ interaction between qubits at idle mode, shown here with PF/I, and entangled mode, shown as PF/E. Switching between I and E modes is reversible operation that requires modulating a circuit parameter to switch. Qubits are undriven at PF/I mode and carry zero residual $ZZ$ interaction.  At PF/E mode, the coupler strength between qubits is changed, and a CR pulse is applied to provide $ZX$-type coupling strength.  This gate combines two essential features: 1) by filtering out parasitic interactions in the presence or absence of external driving, high state fidelity can be achieved on both modes, and 2) it safely provides a faster as well as higher fidelity 2-qubit gate by combining circuit tunability with external driving.

\begin{figure}[t]
	\centering			
\hspace{0.25in}\includegraphics[width=0.35\textwidth]{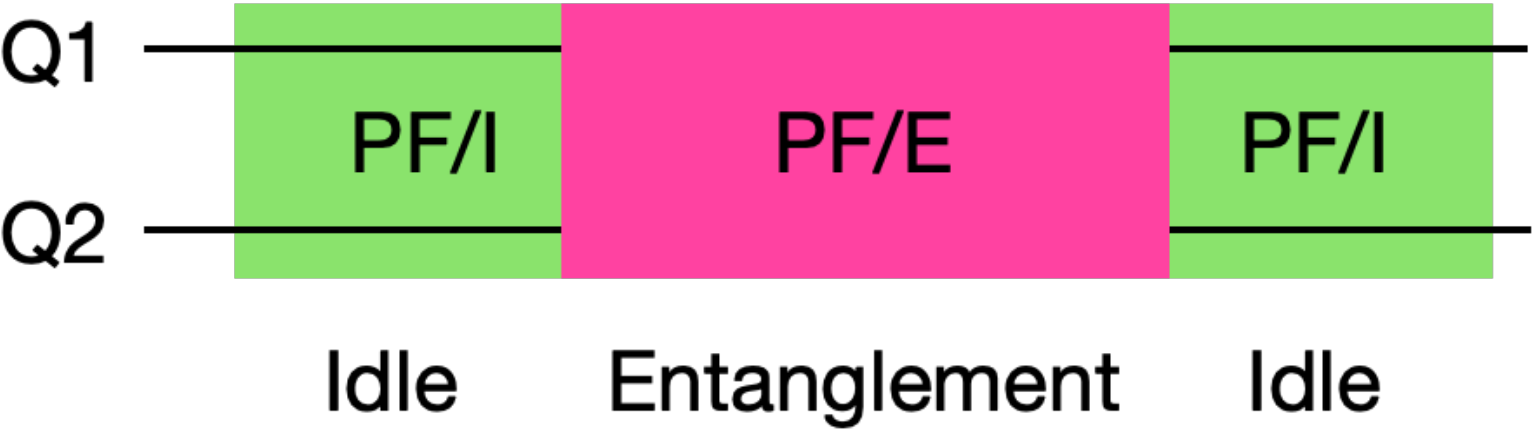}\put(-220.5,48){(a)}\\ \vspace{-0.2in}
	\includegraphics[width=0.42\textwidth]{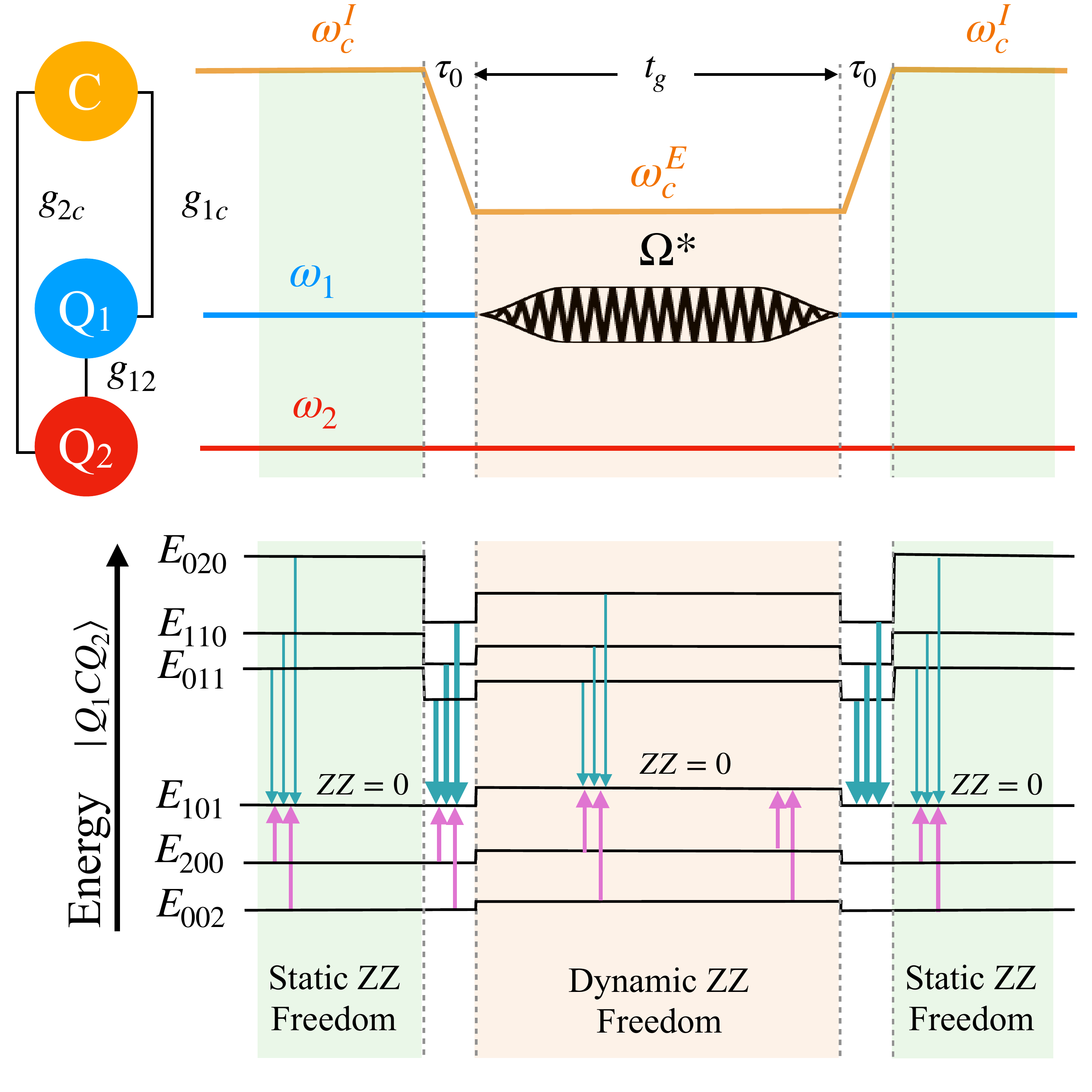}\put(-230,200){(b)}\put(-230,105){(c)}\\
	\vspace{-0.15in}
	\caption{(a) The PF gate at idle mode is active during PF/I operation box (green) and at the entangled mode during the PF/E operation box (pink).  (b) Left: PF gate circuit; Right: PF gate timing and components. $t_g$ is the duration of the entangled mode with $\tau_0$ being the rise/fall time from idle to entangled modes; (c) The energy diagrams at idle and entangled modes -- vertical arrows show level repulsions.} \label{fig:cird}
\end{figure}

 Figure~\ref{fig:cird}(a) shows the schematic of the PF gate that switches qubits Q1 and Q2.
The gate principles are universal for all types of qubits and both harmonic and anharmonic couplers. However, there is a practical preference to use a tunable coupler rather than tunable qubits, since the latter is proven to suffer slightly from relatively lower coherence times that may degrade gate performance~\cite{hutchings2017tunable}.

Let us consider a circuit with qubits Q1 and Q2 coupled via the coupler C and denote their quantum states as $|Q_1, C, Q_2\rangle $.  Schematic circuits can be seen in Fig.~\ref{fig:cird}(b), where Q1 and Q2 interact directly by $g_{12}$ and indirectly by individual couplings to C with coupling strengths $g_{1c}$ and $g_{2c}$.  In principle, qubit and coupler Hamiltonians are similar since  coupler can be considered as a third qubit, i.e. $H_i=\omega_i (n_i) \hat{a}_i^\dagger \hat{a}_i+\delta_i  \hat{a}_i^\dagger \hat{a}_i^\dagger \hat{a}_i \hat{a}_i /2$, with $\hat{a}_i$ ($\hat{a}^\dagger_i$) being annihilation (creation) operator, $\omega_i$  frequency, $\delta_i$ anharmonicity, and  $i=1,2,c$. We can write circuit Hamiltonian as  $H=\sum_{i=1,2,c} H_i + \sum_{i\neq j} g_{ij} (\hat{a}^\dagger_i+\hat{a}_i)(\hat{a}_j^\dagger+\hat{a}_j)$ with $g_{ij}$ being coupling strengths. In the situation where qubits are far detuned from the coupler, $|\omega_{1/2}-\omega_c|\gg |g|$, namely the dispersive regime,  the total Hamiltonian can be perturbatively diagonalized in a higher order of $g/|\omega_{1/2}-\omega_c|$. However, it is important to emphasize that quantum processors can operate beyond the dispersive regime, see Ref.~\cite{ansari2019superconducting}. 

By summing over coupler states and transforming the Hamiltonian into a block diagonal frame \cite{bravyi2011schrieffer--wolff}, one can simplify it as an effective Hamiltonian in the computational Hilbert space of two qubits \cite{magesan2020effective}. This simplification reveals that the two qubits interact only by a $ZZ$ interaction, usually considered unwanted and always on as long as energy levels are not shifted. The computational part of effective Hamiltonian in its eigenbasis, namely `dressed basis', is 
\beq
H_{\rm eff}=-\tilde{\omega}_1 \hat{Z}\hat{I}/2- \tilde{\omega}_2 \hat{I}\hat{Z}/2 + \zeta_s \hat{Z}\hat{Z}/4
\label{eq. Heff} 
\eeq 
with $\tilde{\omega}_i$ being qubit frequency in tilde dressed basis and  $\zeta_s =\tilde{E}_{11}-\tilde{E}_{01}-\tilde{E}_{10}+\tilde{E}_{00}$ being the static level repulsion coefficient in the absence of external driving.  As long as qubits are not externally driven Eq.~(\ref{eq. Heff}) describes the circuit quantum electronics to acceptable accuracy.

 Microwave-driving Q1, namely the `{control}' qubit, at the frequency of Q2, namely the `{target}' qubit, introduces wanted and unwanted transitions between low-lying energy levels. In the bare basis of noninteracting qubits the driving Hamiltonian can be written as  $H_{d} = \Omega \cos (\tilde{\omega}_2 t) (\hat{a}_1^\dagger + \hat{a}_1)$. In the interacting qubit reference frame co-rotating with a driving pulse, the transitions in the leading $\Omega$ order can be divided into two parts: 
\beq 
\label{eq. Hdtot}
H_d=\Hdc+\Hdnc
\eeq 
in which $\Hdc$ ($\Hdnc$) denotes the following computational (non-computational) transitions:
 \beqr
 \label{eq. Horig}
  \Hdc/\Omega && = \lambda_1   \left(  |000\rangle \langle 001| -  |100\rangle \langle 101|\right)  \nonumber  \\ 
  &&+ \lambda_2     \left(  |000\rangle \langle 010| -   |100\rangle \langle 110| \right) + H.c. 
 \eeqr
 and beyond the computational subspace:
 \beqr
 \label{eq. Horigs}
 \Hdnc/\Omega  = && \lambda_3     \left(  |001\rangle \langle 002| -   |101\rangle \langle 102| \right)     \nonumber \\ &&  
 +  |001\rangle \left(     \langle 011| \lambda_4 +  \langle 200| \lambda_5 \right)  \nonumber \\ &&
 +  |010\rangle \left(    \langle 200| \lambda_6 +   \langle 011| \lambda_7  +   \langle 020| \lambda_8 \right)      \nonumber \\ &&
 + \left( \lambda_{9}    | 011\rangle + \lambda_{10}  \Omega |200\rangle + \lambda_{11}   | 002\rangle \right) \langle201|  \nonumber \\&& + H.c.  
 \eeqr 

In the Hamiltonian~(\ref{eq. Horig}) the single-qubit rotations have been dropped out, such as $ |000\rangle \langle 001|$, $|100\rangle \langle 101|$,  $|001\rangle \langle 002|$, and $|101\rangle \langle 102|$. It is important to mention that applying the cross-resonance pulse produces parasitic classical crosstalk between the two qubits, which adds other stray couplings. Some of these terms are eliminated by applying a secondary and simultaneous microwave pulse on the target qubit to drive it at particular amplitude and phase. Additionally, echoing all pulses, or applying a virtual $Z$ gate on target, helps eliminate the $ZI$ component. 

\begin{figure}[h!]
	\centering
	\includegraphics[width=0.4\textwidth]{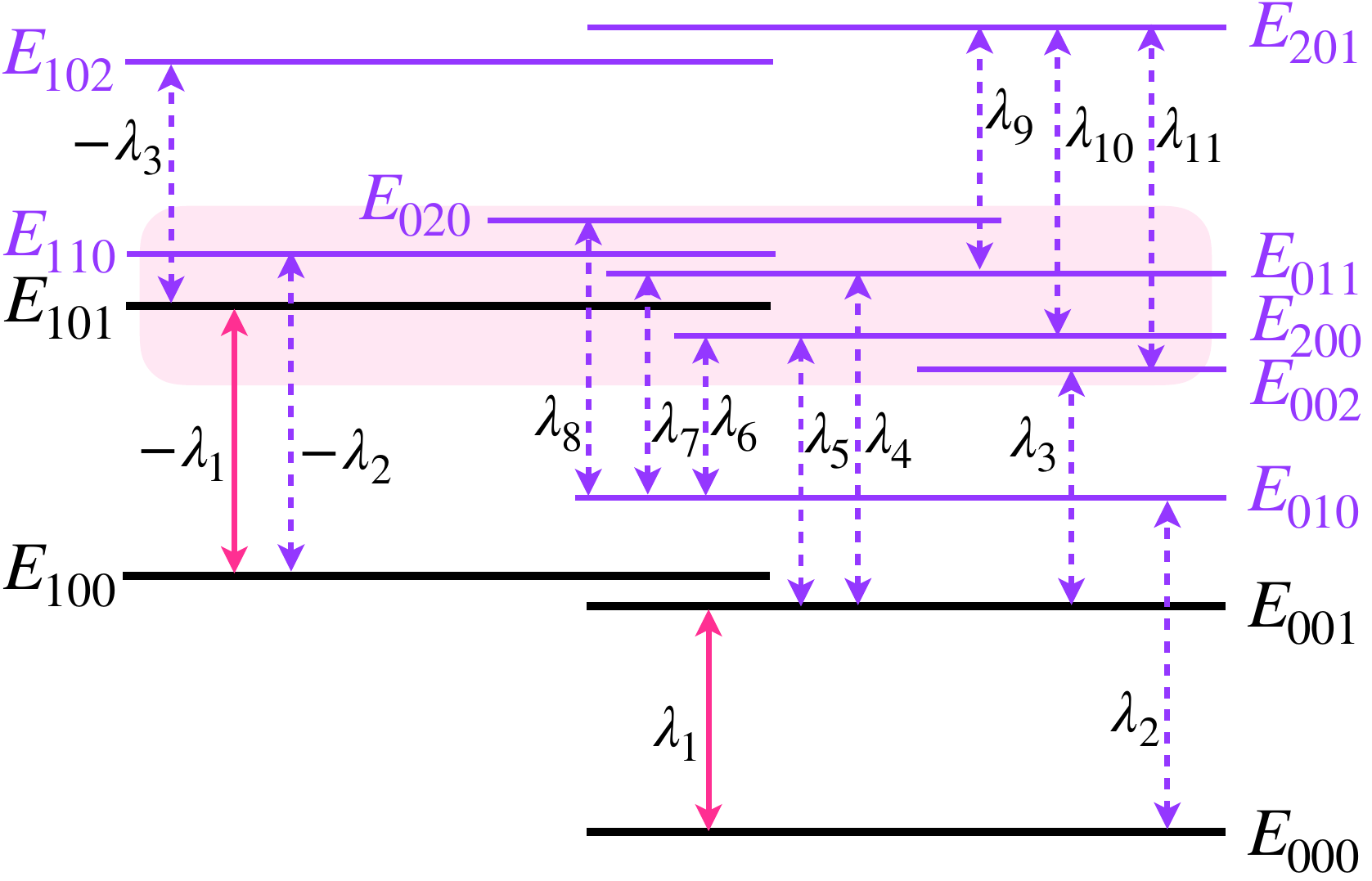}
	\vspace{-0.1in}
	\caption{Energy levels $E_{q_1,c,q_2}$ and microwave-driven transitions in a frame co-rotating with the microwave pulse.  Double arrowed solid (dashed) lines show computational (noncomputational) transitions. The shaded area shows near $E_{101}$ zone.}\label{fig:diag}
\end{figure}

Figure~\ref{fig:diag} shows the computational (non-computational) transitions by solid (dashed) arrows.   Appendix \ref{app.d26} we briefly explain how one can derive this Hamiltonian~(\ref{eq. Horig}) and evaluate all $\lambda$'s in the leading $g^2$ order.

Evidently, in the dispersive regime  by mapping the Hamiltonian (\ref{eq. Horig}) on computational subspace, one can obtain a microwave-assisted part for $ZZ$ interaction between qubits~\cite{magesan2020effective,malekakhlagh2020first-principles}, denoted here by $\zeta_{d}$.  This indicates that the total $ZZ$ interaction in the presence of a driving pulse is $\zeta=\zeta_s + \zeta_d$. 

Let us now supply further details about the static part. The coupler between two qubits can be a harmonic oscillator, such as a resonator, or another qubit with finite anharmonicity. The perturbative analysis of a  harmonic coupler shows that it supplies the effective $ZZ$ coupling $\zeta_{\rm s1}$ between two qubits, which depends on circuit parameters as shown in Eq. (\ref{eq.zeta}). A finite anharmonicity $\delta_c$ for the coupler will add the correction $\zeta_{\rm s2}$.  In Eq.~(\ref{eq.zeta}) we shows both parts in $O(g^4)$:
 \begin{eqnarray}
 \zeta&=&\zeta_{\rm s1}+\zeta_{\rm s2}+\zeta_{d}, \nonumber \\
 \zeta_{\rm s1}& =& \frac{ 2g_{\rm eff}^2\left( \delta_1+\delta_2\right) }{(\Delta_{12}-\delta_2)(\Delta_{12}+\delta_1)}, \nonumber \\
 \zeta_{\rm s2}&=&\frac{8(g_{\rm eff}-\chi g_{12})(g_{\rm eff}- g_{12})}{\Delta_1+\Delta_2-\delta_c}\label{eq.zeta},
\end{eqnarray} 
with $\Delta_{12}=\omega_1-\omega_2$, $\Delta_q=\omega_q-\omega_c$, $\chi=\delta_c/(\Delta_1+\Delta_2)$, and the effective coupling between two qubits is 
\begin{equation}\label{eq.geff}
g_{\rm eff}=g_{12}+\frac{g_{1c}g_{2c}}{2}\sum_{q=1,2}\left(\frac{1}{\Delta_q}-\frac{1}{\Sigma_q}\right),
\end{equation}
with $\Sigma_q=\omega_q+\omega_c$.

At the entangled mode of the PF gate, firstly qubits are coupled by changing circuit parameters; therefore, a non-zero static $ZZ$ interaction is expected to show up between qubits. A cross-resonance pulse is then assisted so that $ZX$ interaction is supplied between qubits. Let us denote the strength of this coupling with $\alpha_{ZX}$.  From Eq.~(\ref{eq. Horig}) one can determine it in the leading perturbative order $\alpha_{ZX} \sim \lambda_1\Omega$ which is in agreement with experiment in weak $\Omega$ regime \cite{sheldon2016procedure}. Any further nonlinearity can be studied in higher orders.  The $ZX$ interaction transforms quantum states by the operator $\hat{U}=\exp (  2\pi i  \alpha_{ZX} \tau  \hat{Z}\hat{X}/2)$ during the time $\tau$ that external driving is active.  In order to perform a typical $\pi/2$ conditional-rotation on the second qubit, i.e. $ZX_{90}$, the two-qubit state must transform by $\exp (i (\pi/2) \hat{Z}\hat{X}/2) $. This indicates that external driving should be switched on for $\tau=1/4\alpha_{ZX}$.  Therefore the stronger $\alpha_{ZX}$ is the shorter time performing the gate takes.  Needless to say, during the whole time the driving is active, this interaction is accompanied by the driving-assisted parasitic interaction $\zeta_{d}$. However, there is a good chance that one can find a large class of parameters at which the total $ZZ$ interaction vanishes. We show in the next section that strengthening $\alpha_{ZX}$ can be found by modulating both coupling strength between qubits and the external driving amplitude. This improves the gate performance by zeroing parasitic interactions and making the gate much faster. 

In the following sections, we will discuss a detailed analysis of several circuit examples and show the performance of the PF gate on them.

\vspace{-0.2in}
\section{Examples of the PF gate} \vspace{-0.1in}\label{sec:realization}

\begin{figure}[h]
	\centering
	\includegraphics[width=0.49\textwidth]{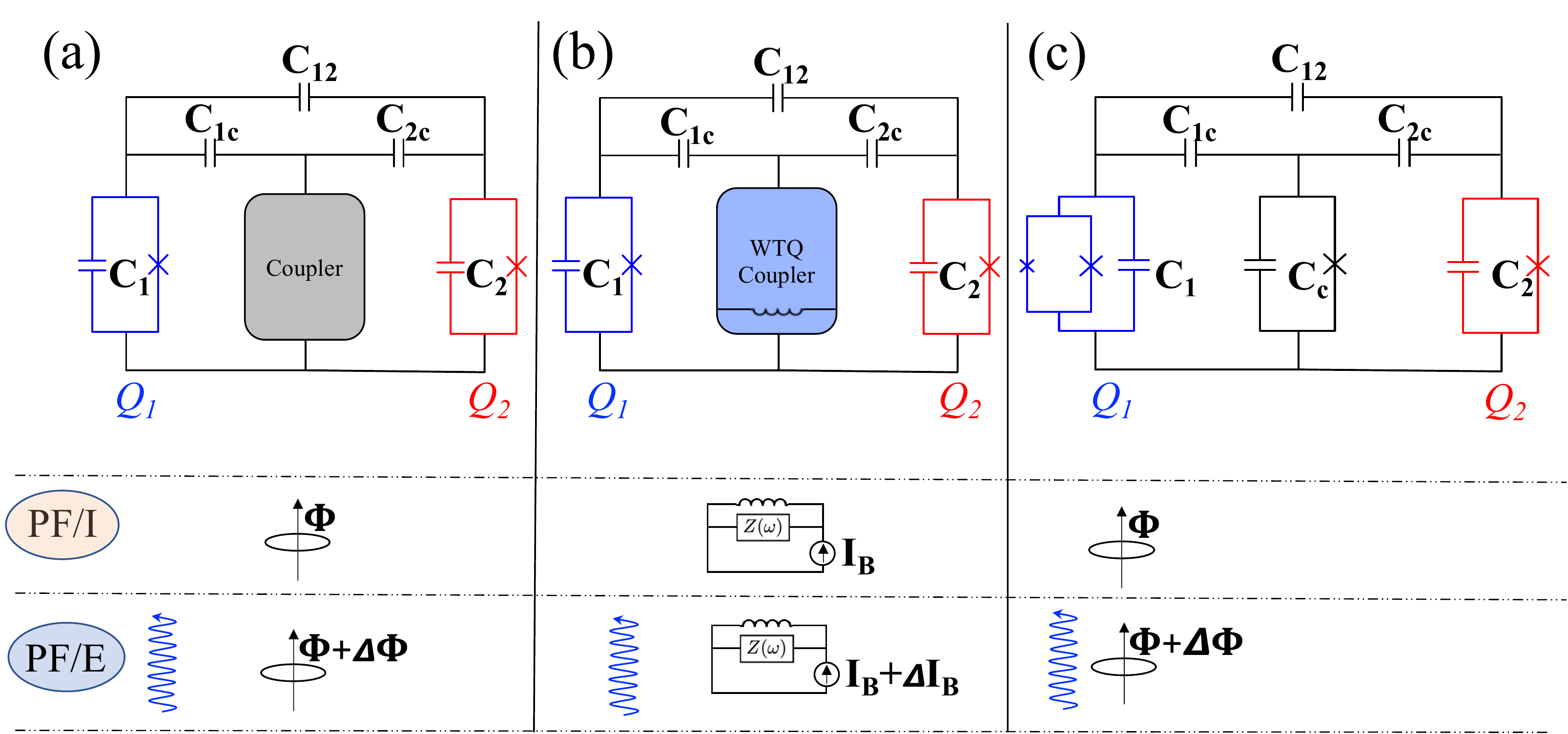}
	\vspace{-0.2in}
	\caption{Three schematic PF-gate circuits, (a) wide frequency-tunable coupler, (b) weakly-tunable coupler, and (c) a tunable frequency qubit. Lower panels show how the PF gate switches between I and E modes.}
	\label{fig:circuit}
\end{figure}

Switching between I and E modes requires a change in circuit parameters before external driving is activated.  In a circuit with two qubits and a coupler there are different possibilities for selecting which is tuned by circuit parameter modulation and which is driven externally. Figure~\ref{fig:circuit} shows three possible examples based on superconducting qubits. In circuit (a) two fixed frequency qubits Q1 and Q2 are coupled by a tunable coupler, which can be another qubit with flux-tunable frequency, so that as one can see in Eq.~(\ref{eq.geff}) changing the flux modulates effective coupling strength between qubits. In circuit (b) two qubits are coupled to a weakly tunable qubit (WTQ), which enables its frequency to be tuned in a small range by manipulating inductive coupling~\cite{chavez-garcia2022weakly}. Circuit (c) is different as it consists of a flux-tunable  Q1 coupled to fixed-frequency Q2 via a fixed-frequency coupler, but it suffers from a relatively limited qubit coherence time. For all three circuits, the E mode can be assisted by externally driving Q1. It is worth mentioning that there are other microwave-activated approaches to implement quantum gates; for instance, imposing additional pulses in the circuit (a) has recently proved useful to execute multiqubit gate experiment~\cite{kim2021highfidelity,baker2021single}.

Here we study circuit (a) with tunability in the coupler, but it does not mean the coupler must be an asymmetric transmon. In this circuit idle mode is obtained by tuning the coupler to the frequency $\wi$, and in entangled mode it is tuned at $\wE$ accompanied by a CR drive, as shown in Fig.~\ref{fig:cird}(b).

To test the performance of the PF gate we numerically study seven sample devices parametrized on the circuit of Fig.~\ref{fig:circuit}(a).  These devices are listed in Table \ref{tab:device}. The direct capacitive coupling $g_{12}$  are grouped in 3 values, the weakest for device 1, intermediate for devices 2, 3, 4, 7, and the strongest for devices 5 and 6. Among devices in the intermediate $g_{12}$ group device 3 has stronger coupler anharmonicity, while devices 2 and 4 have similar coupler anharmonicity, yet in device 4 qubit anharmonicity is stronger. Specifically, device 7 stays out of straddling regime where $|\Delta|>|\delta|$ and the tunable coupler has positive anharmonicity.
In the group of devices 5 and 6 with the strongest $g_{12}$, the qubit-qubit detuning frequency is stronger compared to all other devices with the difference that device 5 is on a hybrid circuit by combining a transmon and a Capacitively Shunted Flux Qubit (CSFQ) while device 6 is on a transmon-transmon circuit. We consider universal qubit-coupler coupling strength $g_{1c}/2\pi=g_{2c}/2\pi=95$~MHz parked at $\omega_c/2\pi=4.8$ GHz for all devices.

\begin{longtable}[t]{@{\extracolsep{\fill}}ccccccc@{}}
	\caption{Device parameters.}
	\label{tab:device}
	\endfirsthead
	\endhead
	\hline\hline
	\centering
	& $\omega_1/2\pi$ &$\omega_2/2\pi$ &$g_{12}/2\pi$ & $\delta_c/2\pi$ & $\delta_1/2\pi$ &$\delta_2/2\pi$ \\ 
	& (GHz) &(GHz) & (MHz) & (MHz) & (MHz) & (MHz)\\
	\hline
	1& 4.25 & 4.20 & 3.76 &$-100$ &$-250$ &$-250$\\
	2 & 4.25& 4.20 & 6.48 &$-100$ &$-250$ &$-250$\\
	3 & 4.25 & 4.20 & 6.48 &$-200$ &$-250$ &$-250$\\
	4 & 4.25 & 4.20 & 6.48 &$-100$ &$-320$ &$-320$\\
	5& 4.00 &4.20&9.48&$-100$&$~~500$&$-250$\\
	6& 4.40 &4.20&9.48&$-100$&$-320$&$-320$\\
	7& 4.50 &4.20&6.48&$+200$&$-250$&$-250$\\
	\hline\hline
\end{longtable}

We evaluate the static $ZZ$ interaction using the parameters listed in Table~\ref{tab:device}. We take three different approaches  for our evaluations. In one approach, we numerically diagonalize the Hamiltonian~(\ref{eq.hcoupler}) in a large Hilbert space. We tested these results with yet another numerical formalism proposed recently in Ref.~\cite{li2021non-perturbative}, namely the Non-Perturbative Analytical Diagonalization (NPAD) method. These two methods give rise to the same result as plotted in Fig. \ref{fig:staticzz} in Appendix \ref{app compare}.  In the same plot we also present the second-order Schrieffer-Wolff perturbative results as SWT, which is consistent with the numerical results only when the coupler frequency is tuned far away from qubits.

\begin{figure}[tp]
	\centering
	\includegraphics[width=0.48\textwidth]{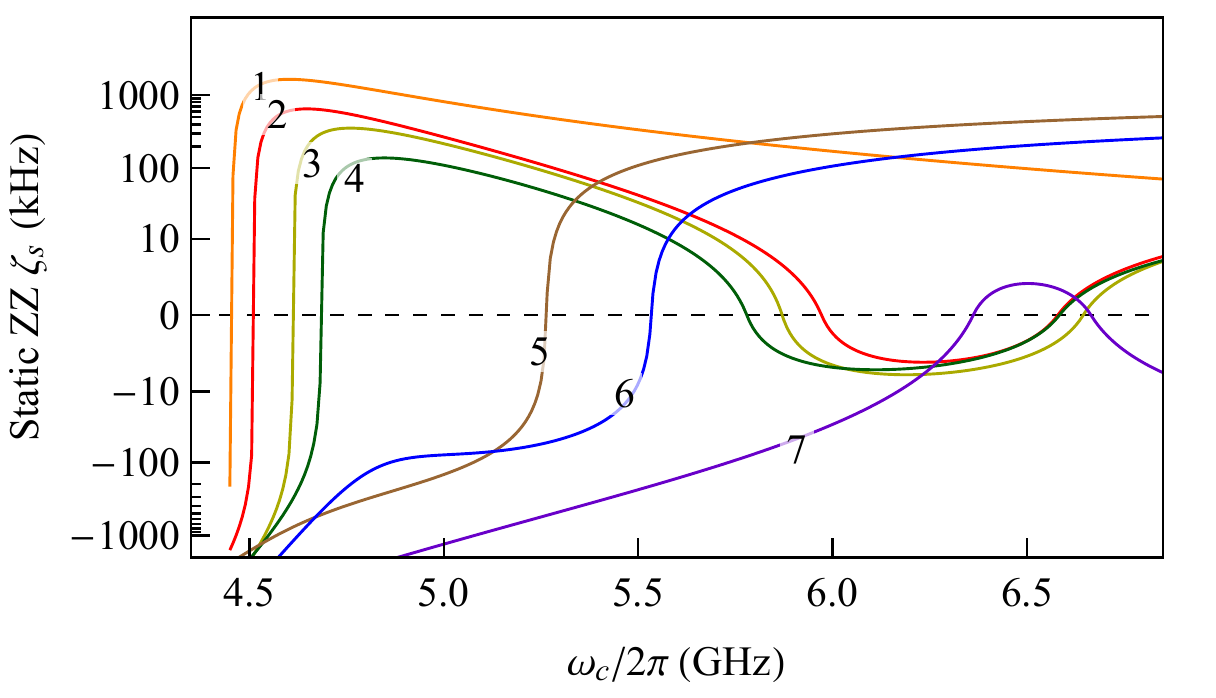}
	\vspace{-0.3in}
	\caption{Numeric static $ZZ$ strength in the seven devices listed in Table~\ref{tab:device} versus coupler frequencies. }
	\label{fig:coupler2zz}
\end{figure}

Figure~\ref{fig:coupler2zz} shows numerical values for the static $ZZ$ interaction at different coupler frequency $\omega_c$ for the seven devices listed in Table \ref{tab:device}.  One can see that all devices possess at least one zero-$ZZ$ point. This will be further discussed below.

\vspace{-0.2in}
\subsection{The idle mode}
\vspace{-0.1in}

In the circuit of Fig.~\ref{fig:circuit}(a), if the coupler frequency is far detuned from qubits, there may exist a particular coupler frequency at which the effective interaction between qubits vanishes, i.e. $g_{\rm eff}=0$. One can easily find the answer using Eq.~(\ref{eq.geff}). However, suppose the coupler frequency is closer to qubits, in that case $\zeta_{s2}$ induced by the anharmonic coupler becomes comparable with $\zeta_{s1}$, making it possible to achieve static $ZZ$ freedom as shown on the far left of Fig.~\ref{fig:coupler2zz}. These two static $ZZ$ freedoms correspond to two types of idle modes; here, we call the far-right point with $g_{\rm eff}=0$ in Fig.~\ref{fig:coupler2zz} as a genuine $ZZ$-free point, and the far left point in Fig.~\ref{fig:coupler2zz} as an affine $ZZ$-free point. There is also a third type of $ZZ$ zeroness which stays between genuine and affine points. However, it always shows up accompanied by at least one of the genuine/affine $ZZ$-free points; we treat it as a trivial solution and will not further discuss it. 

\paragraph*{Genuine idle (GI) mode:} Let us first study the genuine $ZZ$-free point with effective coupling $g_{\rm eff}=0$. As discussed in Ref.~\cite{yan2018tunable} in a circuit couplings between interacting elements are frequency dependent. The qubit-coupler interaction strengths $g_{1c}$ and $g_{2c}$, denoted in Fig.~\ref{fig:cird}(b), can be rewritten in terms of capacitances shown in the analogue circuit of Fig.~\ref{fig:circuit}(a). The  relation between two sets of parameters can be approximated as follows:   $g_{ic}\approx \alpha_{i}  \sqrt{\omega_i \omega_c}$ and $g_{12}\approx \alpha_{12} \sqrt{\omega_1 \omega_2}$ for the qubit label $i=1,2$, with $\alpha_{i}=C_{ic}/2\sqrt{C_i C_c}$ and $\alpha_{12}=(C_{12}+C_{1c} C_{2c} / C_{c}) /2\sqrt{C_1 C_2}$, more accurate derivation can be found in Refs.~\cite{didier2018analytical,ku2020suppression}. By substituting these relations into Eq.~(\ref{eq.geff})  one can find the so-called genuine idle coupler frequency $\wia$ at which qubits are effectively decoupled:
\begin{equation}\label{eq.offwa}
\wia= \frac{\omega_1+\omega_2}{2\sqrt{1-2\alpha_1\alpha_2/\alpha_{12}}}.
\end{equation}

\begin{longtable}{@{\extracolsep{\fill}}cccccccc@{}}
	\caption{Numeric and perturbative coupler frequency $\wia$.}
	\label{tab:offa}
	\endfirsthead
	\endhead
	\hline\hline\\[-2ex]
	\centering
	
	$\wia/2\pi$ (GHz)&1 & 2& 3&4 &5 &6 & 7\\
		\hline\\[-1.5ex]
	Numeric & NA&6.577 &6.643 &6.577&5.261 & 5.532& 6.674 \\[0.5ex]
	Eq.~(\ref{eq.offwa}) & NA &6.522 &6.522 &6.522 &5.278&5.536&6.715\\[0.5ex]
	\hline\hline
\end{longtable}
By substituting $\wia$ in Eq.~(\ref{eq.zeta}) the static $ZZ$ interaction turns out to have a small offset $8g_{12}^2\delta_c/(\omega_1+\omega_2-2\wia)^2$. Usually, this offset in the dispersive regime is only a few kilohertz due to the inaccuracy of the second-order perturbation theory used to derive Eq.~(\ref{eq.zeta}). For example, in the circuit used in Ref.~\cite{sung2021realization} $\alpha_{1/2}\sim10\alpha_{12}\sim0.02$ and $\omega_{1/2}/2\pi\sim4$ GHz, the offset is found approximately $-2$ kHz. This is why we do not limit our analysis in the rest of the paper to perturbation theory. Instead, we take a more accurate approach to numerical Hamiltonian diagonalization. Further comparison can be found in Appendix~\ref{app compare}. The numerical result shows that $\wia$ is slightly shifted from Eq.~(\ref{eq.offwa}) by a few MHz. This difference for the seven circuits is given in Table~\ref{tab:offa}. 

\paragraph*{Affine idle (AI) mode:} When the coupler frequency is closer to qubits, effective coupling $g_{\rm eff}$ is strengthened such that $g_{\rm eff}\gg g_{12}$. In this case by solving $\zeta_{s}=0$ we have the following perturbative idle coupler frequency:
\begin{equation}\label{eq.offwb}
\wib \approx \frac{\omega_1+\omega_2-\delta_c}{2}-\frac{2(\Delta_{12}-\delta_2)(\Delta_{12}+\delta_1)}{\delta_1+\delta_2}
\end{equation}

Table~\ref{tab:offb} compares the numeric simulation of $\wib$ with the perturbative results; note that the perturbative solution beyond the dispersive regime has been ignored.  
\begin{longtable}{@{\extracolsep{\fill}}cccccccc@{}}
	\caption{Numeric and perturbative coupler frequency $\wib$.}
	\label{tab:offb}
	\endfirsthead
	\endhead
	\hline\hline\\[-2ex]
	\centering
	
	$\wib/2\pi$ (GHz)&1 & 2& 3&4 &5 &6 & 7\\
		\hline\\[-1.5ex]
	Numeric & 4.451&4.509 &4.610 &4.683&NA & NA& NA \\[0.5ex]
	Eq.~(\ref{eq.offwb}) & 4.515 &4.515&4.565 &4.587&NA&NA&NA\\[0.5ex]
	\hline\hline
\end{longtable}

\vspace{-0.2in}
\subsection{The entangled mode}\label{gmode}
\vspace{-0.1in}
The PF gate is switched on the entangled mode in two steps: in the first step, coupler frequency is brought out of the $\wi$ value so that static $ZZ$ becomes nonzero. In the second step, qubits are driven externally by microwave pulse with amplitude $\Omega$. 

One two-qubit circuit example is to consider that we use a microwave pulse to drive the `control' qubit, i.e. the qubit whose second excited level is higher, with the frequency of the other qubit, namely the `target' qubit. As discussed above, this driving is called cross resonance drive and is supposed to supply only the conditional $ZX$-type interaction between qubits. However, the desired external force is accompanied by 3 types of unwanted operators  listed below:
\begin{enumerate}
	\item Classical crosstalk: supposedly triggered by the classical translation of driving electromagnetic waves to the position of target qubit,
	\item Control $Z$ rotation: due to driving control qubit with  qubits detuning frequency,  
	\item Microwave-assisted $ZZ$ interaction: triggered by $\Omega$-transition between computational and noncomputational levels ---listed in Eq. (\ref{eq. Horig}) and shown in  Fig.~\ref{fig:diag} --- that changes $\zeta$ level repulsion.  
\end{enumerate}

The first two can be eliminated as described in Ref. \cite{mckay2017efficient}: driving target qubit with a second pulse to eliminate classical crosstalk and either echoing the pulses or software-counterrotating control qubit to eliminate its detuning $Z$ rotation. However, these methods do not eliminate the third one. To eliminate it, we proposed a method called `dynamic freedom',  which sets total $ZZ$ to zero by fine-tuning microwave parameters to cancel out the static parasitic interaction~\cite{xu2021zz-freedom}. The PF gate takes advantage of the dynamic freedom in the entangled mode by combining microwave driving with a tunable coupler.

Let us recall that after eliminating the classical crosstalk and control $Z$ rotation,  external driving activates the Hamiltonian (\ref{eq. Horig}) with transitions within and outside of computational levels shown in Fig.~\ref{fig:diag}.  By block-diagonalizing the Hamiltonian to the computational subspace one can find the following simplified version:  
\beq
H_{d}(\Omega)=\alpha_{ZX} (\Omega) \hat{Z}\hat{X}/{2}+\zeta_d (\Omega) \hat{Z}\hat{Z}/{4}
\label{eq.zx}
\eeq

Perturbation theory  helps determine $\zeta_d$ and $\alpha_{ZX}$ in terms of driving amplitude $\Omega$. Results show that  $\alpha_{ZX}(\Omega)$ depends linearly on $\Omega$ in the leading order and $\zeta_d(\Omega)$ depends on $\Omega^2$ (For details see Eq.~(15) and Fig.~5,6 in \cite{xu2021zz-freedom}).    One may expect that higher order corrections can be worked out by adding terms with larger natural number exponent; however, comparing results with experiments has shown in the past that perturbation theory is not accurate beyond leading order \cite{magesan2020effective}. Alternatively we use a nonperturbative approach, the so-called Least Action (LA)~\cite{cederbaum1989block,magesan2020effective}.

\begin{figure*}[t]
\hspace{0.3in}\includegraphics[width=0.7\textwidth]{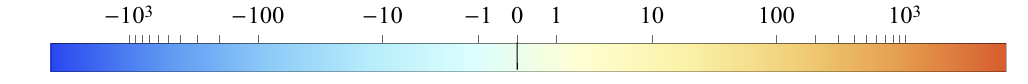}\put(-205,30){Total $ZZ$  (kHz)}\\
\includegraphics[width=0.4\textwidth]{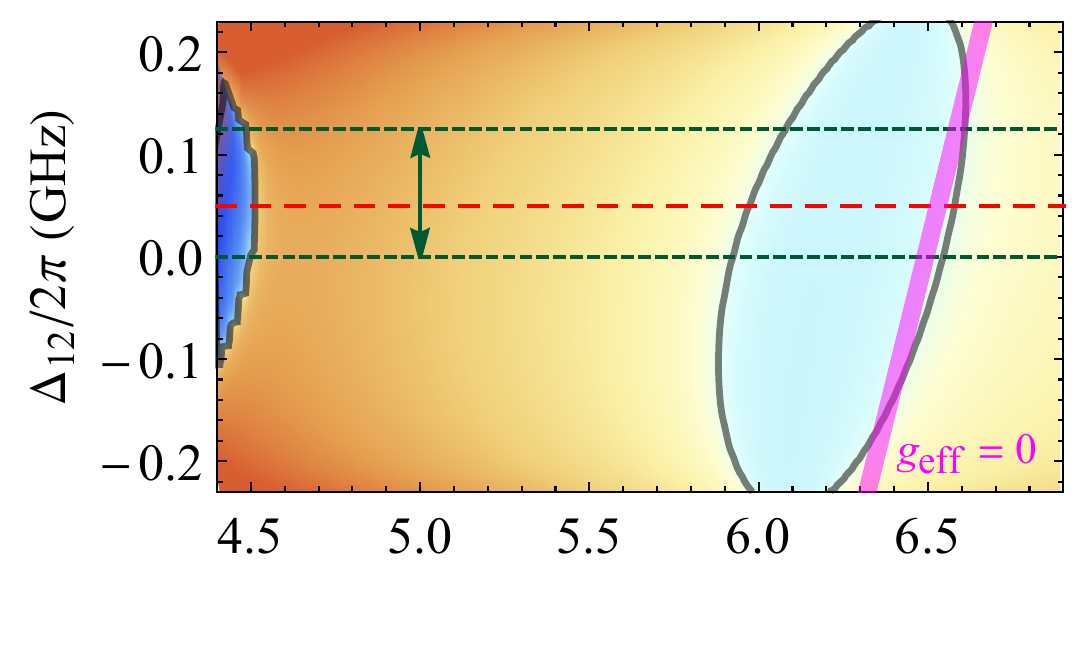}\hspace{-0.2in}
\includegraphics[width=0.4\textwidth]{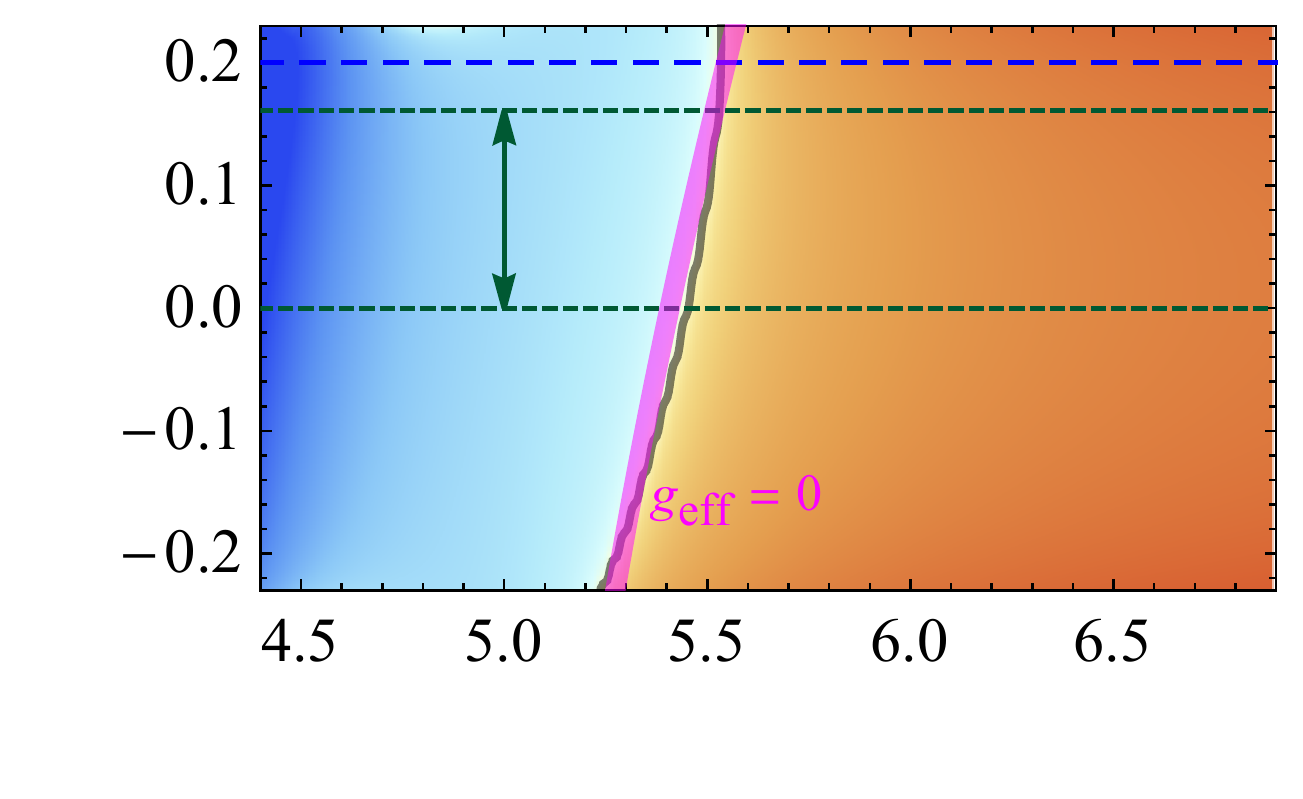}
\put(-387,118){(a)}\put(-195,118){(b)}
\put(-240,84){2}\put(-80,112){6}
\put(-320,38){$\Omega$=0}\put(-65,38){$\Omega$=0}
\put(-315,91){\textcolor{phthalogreen}{$|\delta_1/2|$}}\put(-122,91){\textcolor{phthalogreen}{$|\delta_1/2|$}}
\put(-233,44){\textcolor{black}{$\ast$}}\put(-217,104){\textcolor{black}{$\ast$}}
\put(-108,40){\textcolor{black}{$\ast$}}\put(-93,110){\textcolor{black}{$\ast$}}\\
\vspace{-0.25in}
\includegraphics[width=0.4\textwidth]{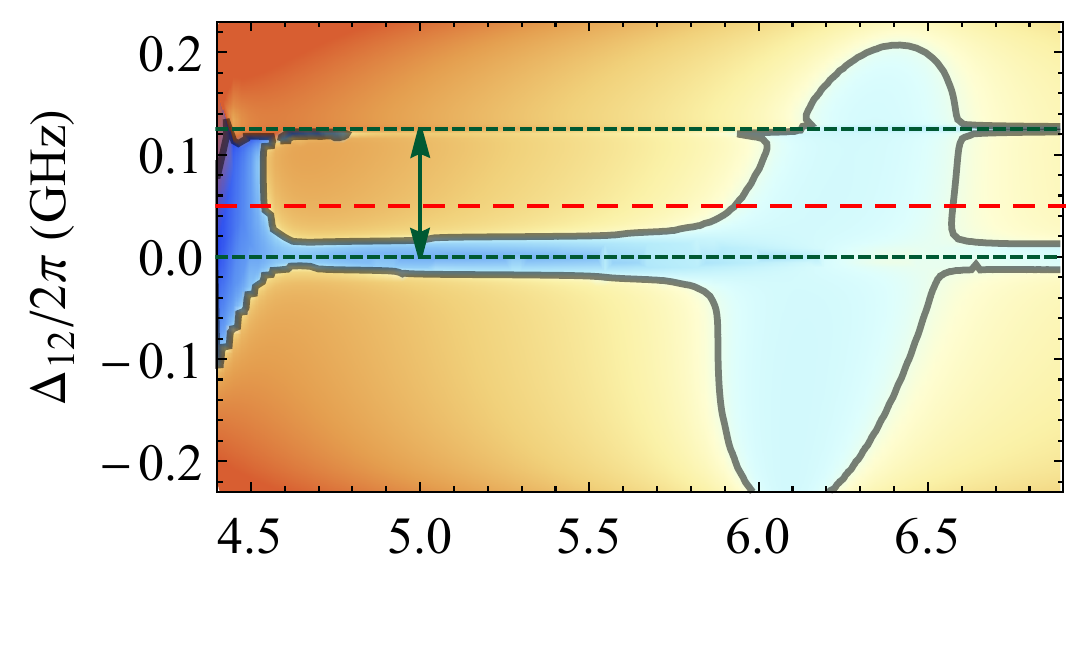}\hspace{-0.2in}
\includegraphics[width=0.4\textwidth]{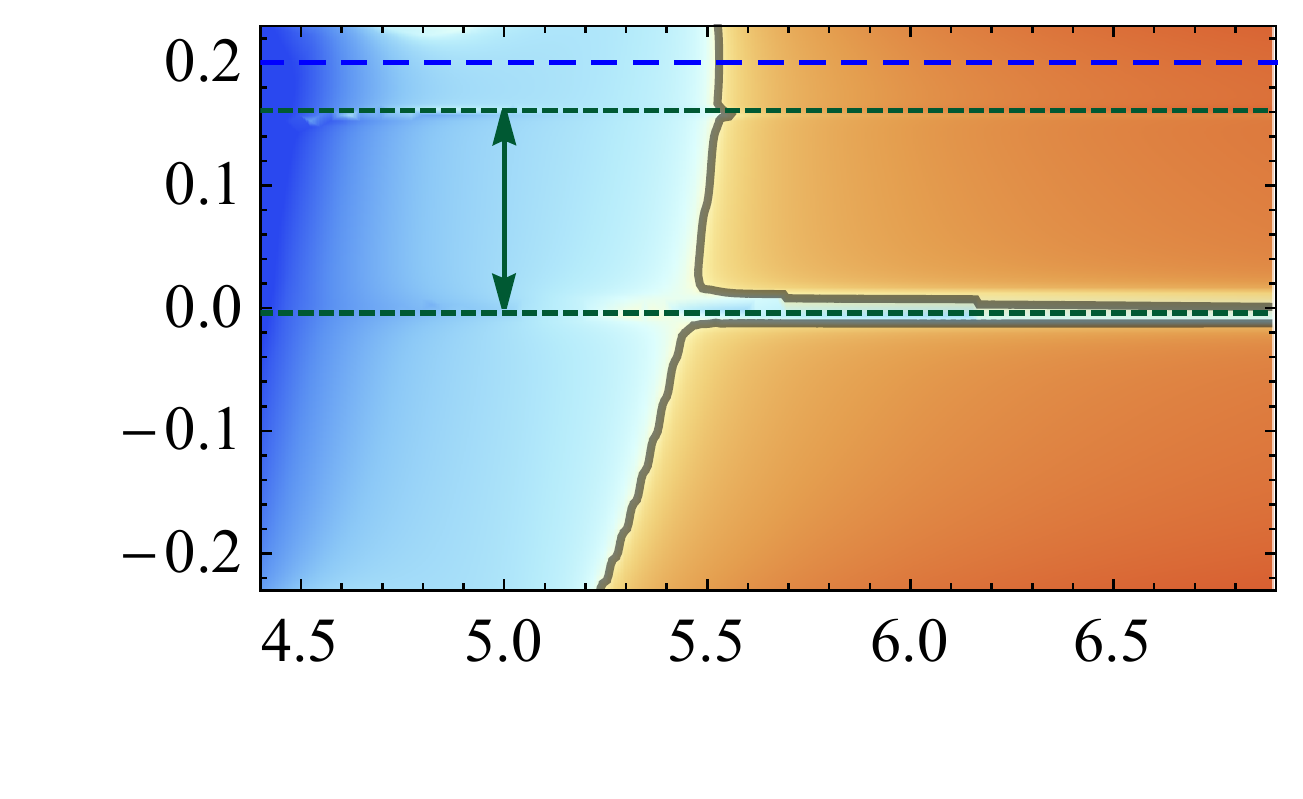}
\put(-387,118){(c)}\put(-195,118){(d)}
\put(-240,84){2}\put(-80,112){6}
\put(-320,38){$\Omega=20$ MHz}\put(-65,38){$\Omega=20$ MHz}
\put(-315,91){\textcolor{phthalogreen}{$|\delta_1/2|$}}\put(-122,91){\textcolor{phthalogreen}{$|\delta_1/2|$}}\\
\vspace{-0.25in}
\includegraphics[width=0.4\textwidth]{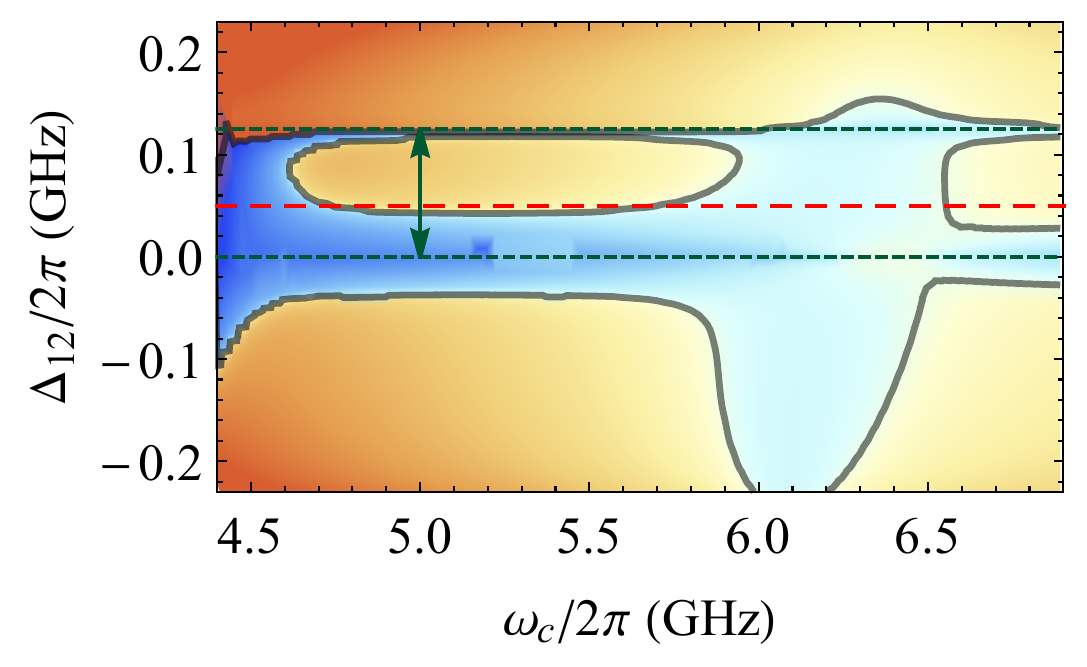}\hspace{-0.2in}
\includegraphics[width=0.4\textwidth]{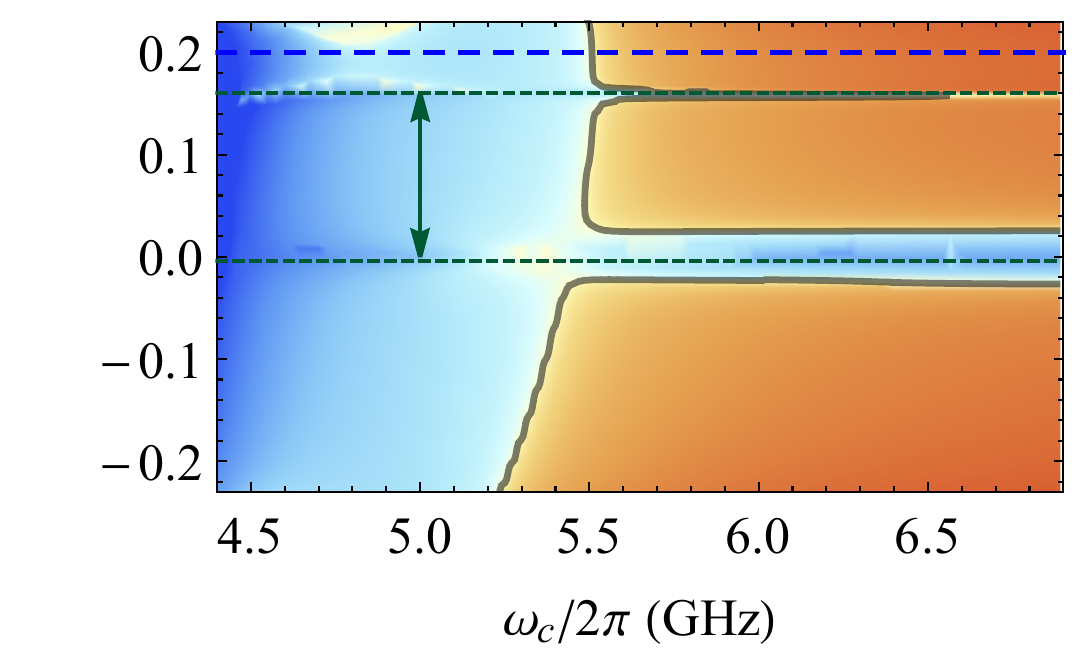}
\put(-387,118){(e)}\put(-195,118){(f)}
\put(-240,84){2}\put(-80,112){6}
\put(-320,38){$\Omega=40$ MHz}\put(-65,38){$\Omega=40$ MHz}
\put(-315,91){\textcolor{phthalogreen}{$|\delta_1/2|$}}\put(-122,91){\textcolor{phthalogreen}{$|\delta_1/2|$}}
  \vspace{-0.13in}
  \caption{Total $ZZ$ interaction as a function of qubits detuning frequency $\Delta_{12}$ and coupler frequency $\omega_c$ with parameters similar to device 2 in (a,c,e) and similar to device 6 in (b,d,f).  $\Omega=0$ in (a,b), $\Omega=20$~MHz in (b,c), and $\Omega=40$~MHz in (e,f). Red (blue) lines denote the labelled devices 2 (devices 6). Black boundaries are $ZZ$-free zones and magenta boundaries are $g_{\rm eff}$-free zones. In particular, in the absence of external microwave driving as shown in (a) and (b), the line of $g_{\rm eff}=0$ intersects static $ZZ$-free borderline in two asterisk points. These points are the genuine points, and all other points on the static $ZZ$-free borderlines are affine.}\label{fig:dy2d}
\end{figure*}
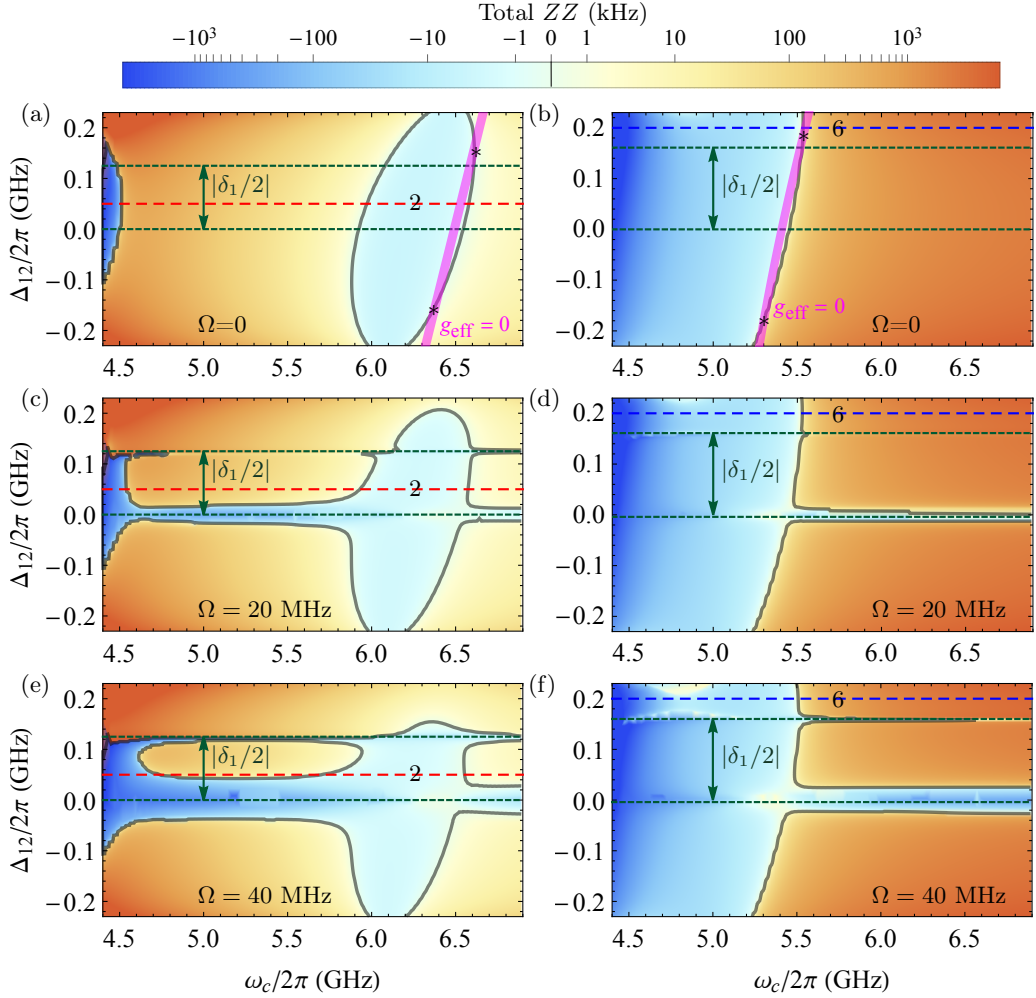

  Our numerical analysis evaluates total parasitic interaction $\zeta$ by adding the driving part to the static part.  We plot the total $ZZ$ interaction in Fig.~\ref{fig:dy2d} in a large range of qubits frequency detuning $\Delta_{12}$ and coupler frequency $\omega_c$ for two sets of circuit parameters. Left (Right) column plots show simulations for a set of parameters similar to device 2 (6) except that here we keep $\Delta_{12}$ variable. On the dashed lines labelled by 2 and  6 the detuning frequencies are fixed to values given in Table~\ref{tab:device}.  We plotted three sets of driving amplitudes in each row: Fig.~\ref{fig:dy2d}(a,b) show no driving $\Omega=0$ to study static level repulsions, Fig. \ref{fig:dy2d}(c,d) shows total $ZZ$ interaction after we apply driving with amplitude $\Omega=20$~MHz, and Fig.~\ref{fig:dy2d}(e,f) doubles the amplitude to $\Omega=40$~MHz.

 In these plots, we show the total parasitic interactions can be either positive (in red), or negative (in blue). The zero $ZZ$ devices are shown in black boundaries between the two regions.  In Fig. (\ref{fig:dy2d}) by closely examining $\zeta$ variation with $\omega_c$ one or more than one zero  points can be found for a device with fixed $\Delta_{12}$. More details with stronger driving amplitude can be found in Appendix~\ref{sec:dzzf}.
 
 In general, there are two types of $ZZ$-free  boundaries: Type I can be found in regions where $\zeta$ values become shallow by gradually being suppressed, and they change the sign smoothly in white areas. Examples are zeros on the closed loop in (a,c,e) and the boundary in the middle of (b,d,f). In type II the $\zeta$ values abruptly change the sign between dark blue and red areas in a narrow domain of parameters. Examples are the far left side boundaries in (a,c,e). These correspond to two types of PF gate: {\it genuine PF gate} starting from genuine idle mode to E mode with type I freedom and {\it affine PF gate} starting from affine idle mode to E mode with type II freedom.

 

 External driving in Fig.~\ref{fig:dy2d}(c-f) leaves a large class of devices with zero total $ZZ$ interaction; however, it is noticed by comparing total $ZZ$ with the static one that external driving distorts the freedom boundaries.  In (a,c,e) subplots, which describe the same devices,  by increasing $\Omega$ the closed-loop surrounding a blue island on the right is shrunk, while a new closed-loop appears in the middle surrounding a red island. These boundaries are additionally distorted for devices with resonant frequency qubits $\Delta_{12}=0$ and devices at the symmetric point with  $\Delta_{12}=-\delta_1/2$. Perturbation theory shows that $\zeta$ diverges at these two points.  We show these points in darker green dashed lines. Our nonperturbative numerical results based on the LA method show $\zeta$ stays finite, however by increasing driving power, near these detuning values microwave-assisted component $\zeta_d$ is largely magnified and heavily dominates total $ZZ$ therefore zero boundaries are largely distorted.  Further discussion about the derivation of microwave-assisted components $\zeta_d$ can be found in Appendix~\ref{app.d26}.

\begin{figure}[tp]
	\centering
	\includegraphics[width=0.48\textwidth]{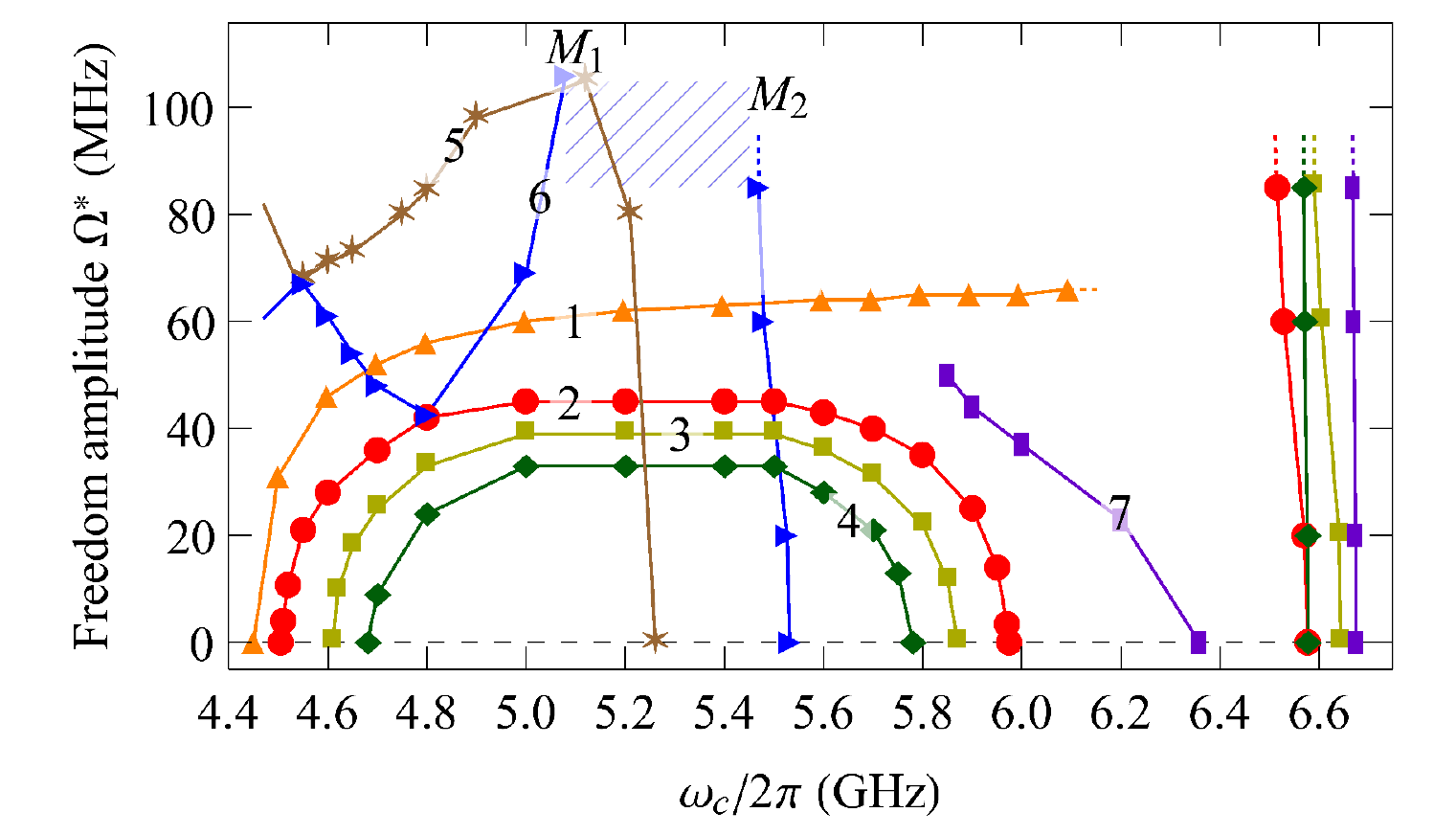}
	\put(-250,135){(a)}\\
	\vspace{-0.in}
	\includegraphics[width=0.48\textwidth]{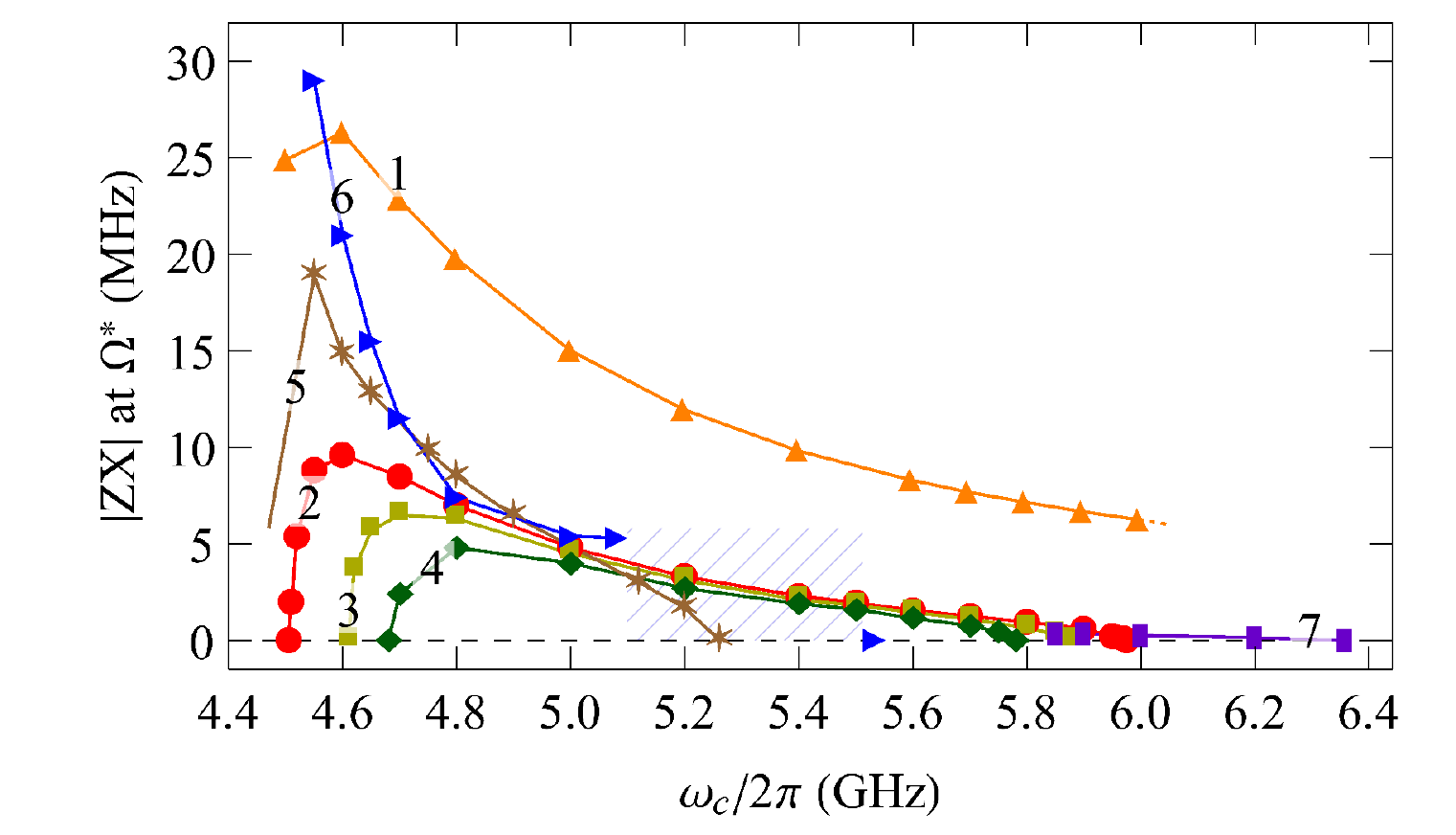}
	\put(-250,135){(b)}
	\vspace{-0.1in}
	\caption{(a) Freedom amplitude $\Omega^*$ as a function of the coupler frequency in devices 1-7.  (b) Corresponding $ZX$ rate. The shaded area between $M_1$ and $M_2$ indicates the absence of dynamic $ZZ$ freedom.}
	\label{fig:coupler2dy}
\end{figure}

Let us study the devices listed in Table \ref{tab:device} and we externally drive each with driving amplitude $\Omega$ and then evaluate the coupler frequency for dynamic freedom. Any amplitude associated with a $ZZ$-free coupler frequency is named the {\it{freedom amplitude}} denoted by $\Omega^*$. Figure~\ref{fig:coupler2dy}(a) shows driving device 1 with amplitudes below 60~MHz sets total parasitic interaction to zero. Devices 2-4 show three such frequencies in a somewhat weaker domain of freedom amplitudes, with an exciting feature on the rightmost one, near $\sim$6.6~GHz. Increasing driving amplitudes does not change the strongest $\omega_c^{\rm E}$, therefore at this frequency, not only static level repulsion $\zeta_s$ is zero, but also driving assisted component $\zeta_d$ vanishes since $g_{\rm eff}=0$. Devices 5 and 6 show decoupling frequencies below a certain driving amplitude, adding more $ZZ$-free coupler frequencies. In device 6, however, there is a frequency domain between $M_1$ and $M_2$ in which parasitic freedom is not expected to take place, and we indicate it with the shaded region and we will come back to it later.  For device 7 staying beyond straddling regime, the static $ZZ$ freedom is only realized at a higher coupler frequency.

A two-qubit gate is not only needed to have high fidelity, but also it must be fast because such gates can perform many operations during qubit coherence times. as discussed above, it is vital that external driving supplies a strong $ZX$ interaction, mainly because the strength of $ZX$ interaction, i.e. $\alpha_{ZX}$ scales inversely with the time that consumes to perform the gate, the so-called gate length $\tau \sim 1/\alpha_{ZX}$. Therefore the entangled mode of the PF gate must be tuned on a coupler frequency that not only is located on a $ZZ$-free boundary but also present a short gate length.  Figure~\ref{fig:coupler2dy}(b) plots $ZX$ strength at all freedom amplitudes and shows that  $ZX$ rate is stronger at lower $\omega_c$'s. Therefore a $ZZ$-free coupler frequency with strong $ZX$ strength can be used for $\omega_c^{\rm PF/E}$. One exception is device 7 which stays out of the straddling regime and has a weaker $ZX$ rate, so it is not feasible to implement a CR-like gate and will not be discussed later.


One of the advantages of our numerical analysis is that it can predict nonlinear correction in both $\zeta$ and $\alpha_{ZX}$ denoted in Eq.~(\ref{eq.zx}).   Perturbation theory determines the leading order of $\alpha_{ZX}$ and $\zeta_d$ are linear and quadratic in $\Omega$, respectively~\cite{xu2021zz-freedom}. The perturbative theory considers higher-order terms with next natural-number exponents above the leading terms; however, compared with the experiment, those results are not accurate beyond leading order~\cite{magesan2020effective}. Since our approach is different, we consider higher-order corrections in real-number exponents: 
\beqr     
\zeta_d (\Omega)&=&\eta_2\Omega^2+\eta_a\Omega^a, \  {\rm with} \ \ a>2,\\  
\alpha_{\rm zx} (\Omega) &=&\mu_1\Omega+\mu_b\Omega^b,\ \  {\rm with }\ \ b>1. 
\label{eq.zz}
\eeqr

\begin{figure}[h!]
	\centering
	\includegraphics[width=0.48\textwidth]{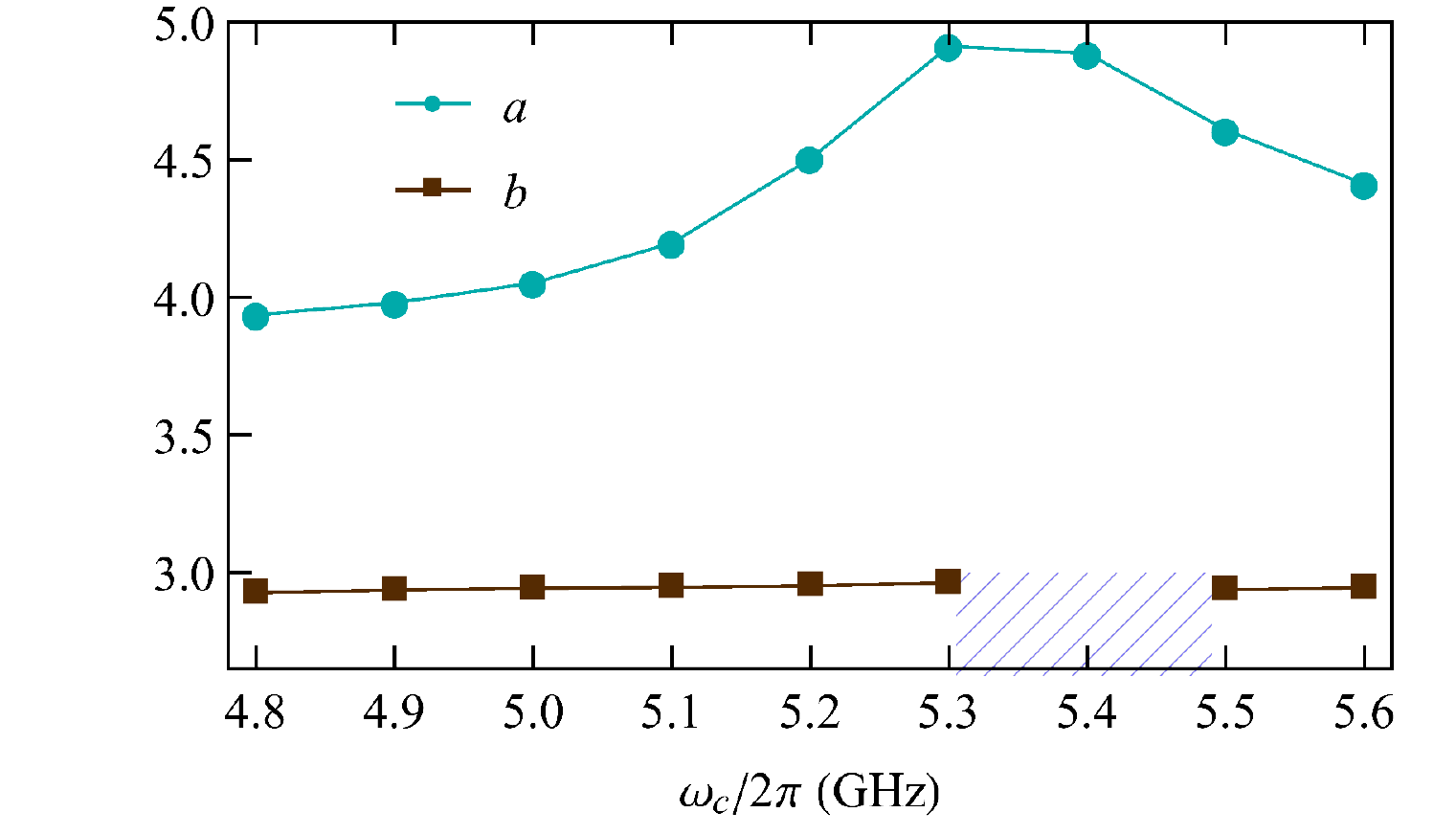}
	\vspace{-0.3in}
	\caption{Beyond 2nd (1st) order exponent of $\Omega^a$  ($\Omega^b$) in $\zeta_d$ ($\alpha_{ZX}$)  at different coupler frequency $\omega_c$ for device 6. The shaded area denotes the region where effective coupling $g_{\rm eff}$ is small and starts to change its sign.}
	\label{fig:horder}
\end{figure}

Our numerical results for $\zeta_d$ in device 6 estimates the exponents $a$ and $b$ at different coupler frequencies $\omega_c$.  The result is summarized in Fig.~\ref{fig:horder}, in which far-left points are similar to perturbative results; i.e. $a=4$ in $\zeta$ and $b\sim 3$ in $\alpha_{ZX}$.   However, there is a domain of frequency in which $a$ increases and nearly reaches 5. Moreover, within the same domain $ZX$ rate vanishes and this makes it meaningless to calculate $b$ exponent in that region. 

In device 6 the higher-order term of $\zeta_d$ has the opposite sign of $\eta_2 \Omega^2 +\zeta_s$ and is the dominant term in the coupler frequency domain 5.1--5.5~GHz. Therefore in this domain total $ZZ$ cannot vanish.  This describes why there is a shaded area in Fig.~\ref{fig:coupler2dy}(a) in device 6 where entangled mode cannot be found. More details can be found in Appendix~\ref{app.higherorder}.

\section{OFF-ON-OFF Error mitigation}\label{sec:error}
\vspace{-0.1in}
Switching from Idle mode to entangled mode takes place in two steps: first coupler frequency is changed,  then the microwave pulse is activated. The other way around needs to take place in the reversed order: first, the microwave pulse is switched off, then the coupler frequency is changed. In each step, there is a possibility that quantum states accumulate errors.  Although the PF gate can effectively eliminate universal $ZZ$ interaction, it still suffers from unwanted transitions during coupler frequency change.  Moreover, limited qubit coherence time can be another source of fidelity loss. Here we quantify the performance of both the genuine PF gate and affine PF gate by calculating the metric of gate fidelity. 

\vspace{-0.2in}
\subsection{Error during coupler frequency variation}
\vspace{-0.1in}
   
For the genuine PF gate, the coupler frequency is far detuned from the frequency of qubits; switching the mode to entangled mode requires that the coupler frequency is brought much closer to the qubits. While for the affine PF gate, the coupler frequency is near qubits, then switching the mode to entangled mode only needs a slight change in the coupler frequency, e.g. with a WTQ whose coherence time is nearly unchanged and comparable with transmons~\cite{chavez-garcia2022weakly}.  Figure~\ref{fig:adia}(a) sketches the two types of PF gate implemented in device 2. In either case, to avoid reinitialization of qubit states after a frequency change, we can perform the frequency change so that leakage does not take place from the computational subspace to other energy levels. This mandates performing the coupler frequency change adiabatically \cite{dicarlo2009demonstration}. 

The leakage rate out of computational levels depends on the ramping speed of coupler frequency $d\omega_c/dt$. In particular, if the coupler frequency is tuned by external magnetic flux $f=\Phi_{\rm ext}/\Phi_0$ with $\Phi_0$ being flux quantum unit, the rate of coupler frequency change can be written in terms of $d f/dt$ \cite{xu2020high-fidelity}.  Here we compare two protocols for the pulse envelopes to quantify the leakage due to $\omega_c$ modulation. Coherence times of qubits and coupler are ideally assumed to be all the same: i.e. $\{T_1^{(1)}, T_1^{(c)},T_1^{(2)}\}=\{T_2^{(1)}, T_2^{(c)},T_2^{(2)}\}=\{200,200,200\}$~$\mu$s. 

\subsubsection{Genuine PF gate} 
A genuine PF gate is a condition for the cross resonance on/off switch with off mode being set for decoupled qubits with $g_{\rm{eff}}=0$, where genuinely direct qubit-qubit coupling cancels out indirect coupler-qubit coupling.

Figure~\ref{fig:adia}(b) shows two pulses that we prepared for being used on device 2: a hyperbolic tangent envelope pulse in a solid line and a flat-top Gaussian envelope in a dashed line.  The qubits are decoupled at $\wia/2\pi=$6.577~GHz. Figure~\ref{fig:coupler2dy}(b) shows that $\alpha_{ZX}$ is rather strong --- nearly $\sim$5~MHz --- at the frequency 4.8~GHz which we take as $\wE/2\pi$.  Note that much stronger $\alpha_{ZX}$ of nearly 10~MHz is also possible on this device which corresponds to a 0.2~GHz smaller coupler frequency. 

\subsubsection{Affine PF gate} 
An affine PF gate is a condition for the cross resonance on/off switch with off mode being set for coupled qubits with a finite $g_{\rm{eff}}$. 

\begin{figure}[tp]
	\centering
	\includegraphics[width=0.45\textwidth]{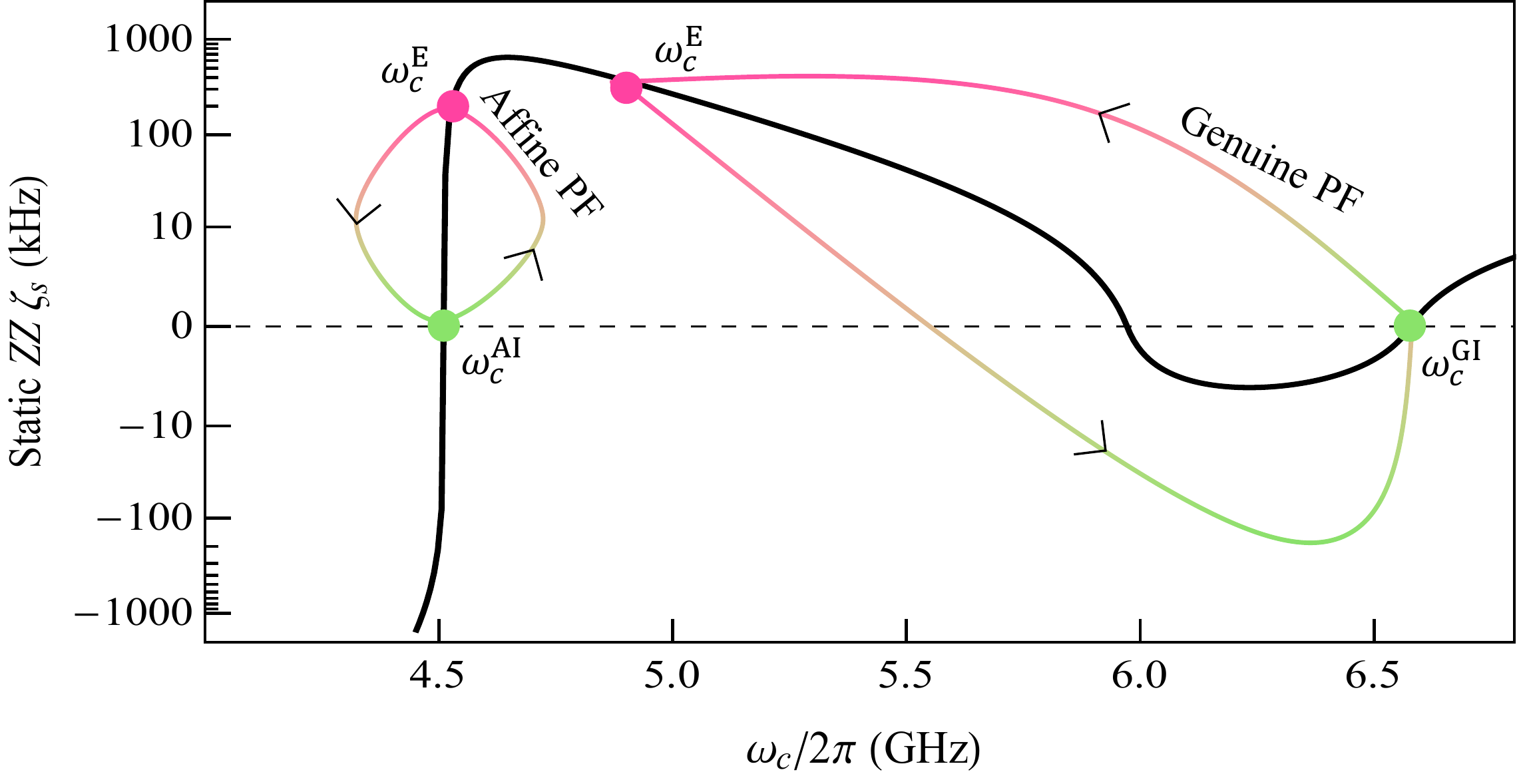}\put(-244,115){(a)}\\
	\vspace{0.02in}
	\hspace{0.05in}\includegraphics[width=0.23\textwidth]{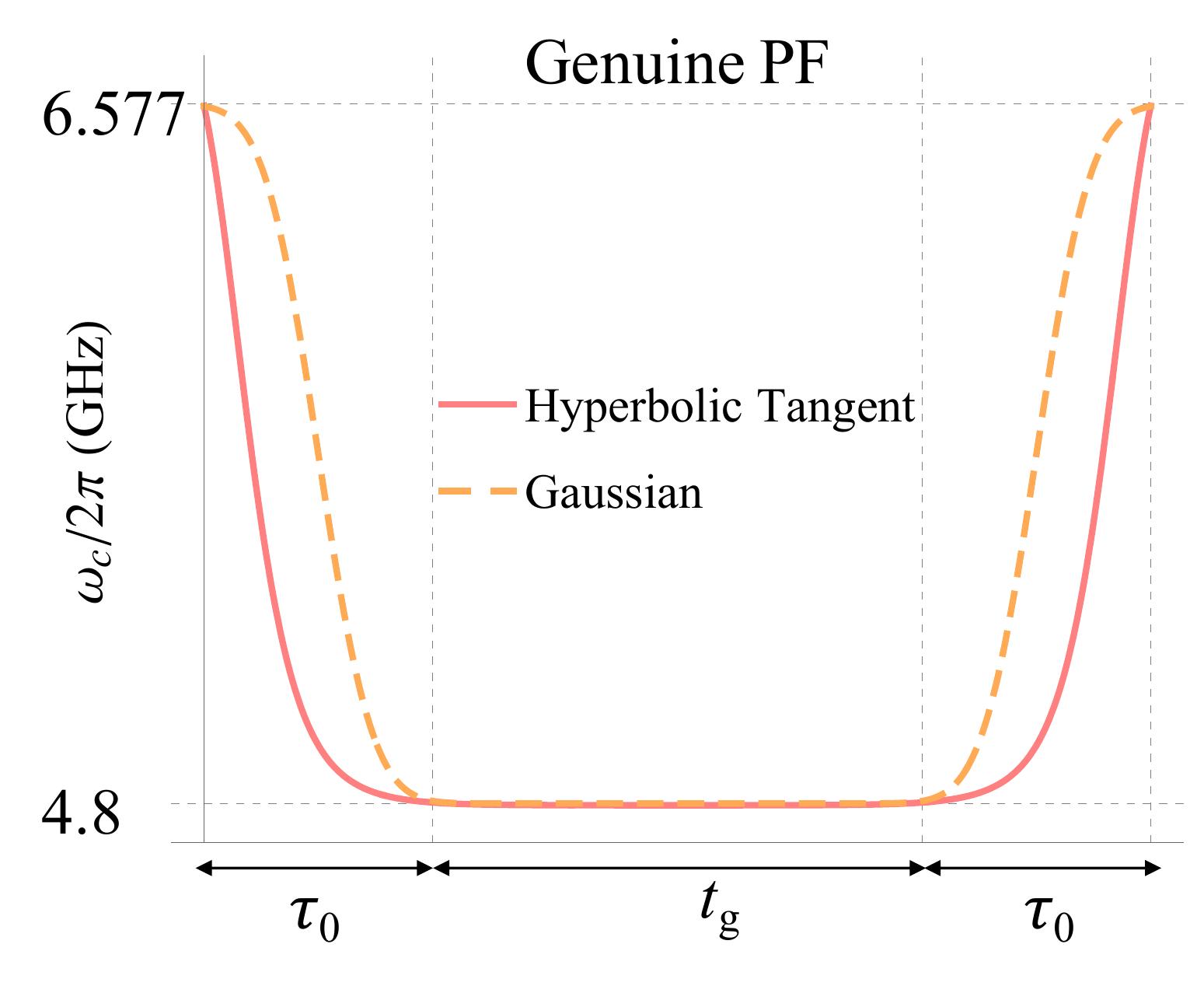}\put(-128,85){(b)}\hspace{0.05in}
	\includegraphics[width=0.23\textwidth]{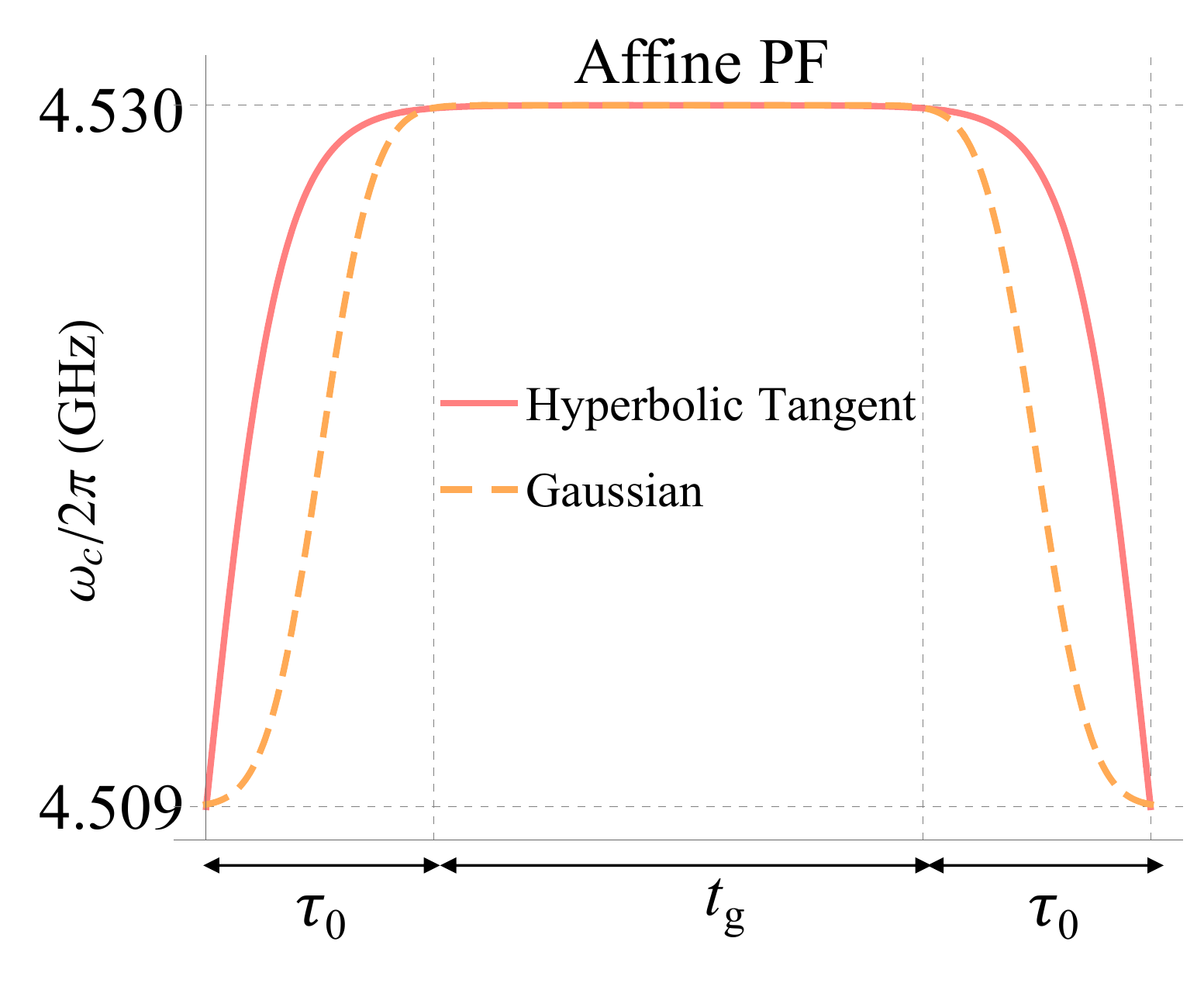}\put(-125,85){(c)}\\
	\includegraphics[width=0.41\textwidth]{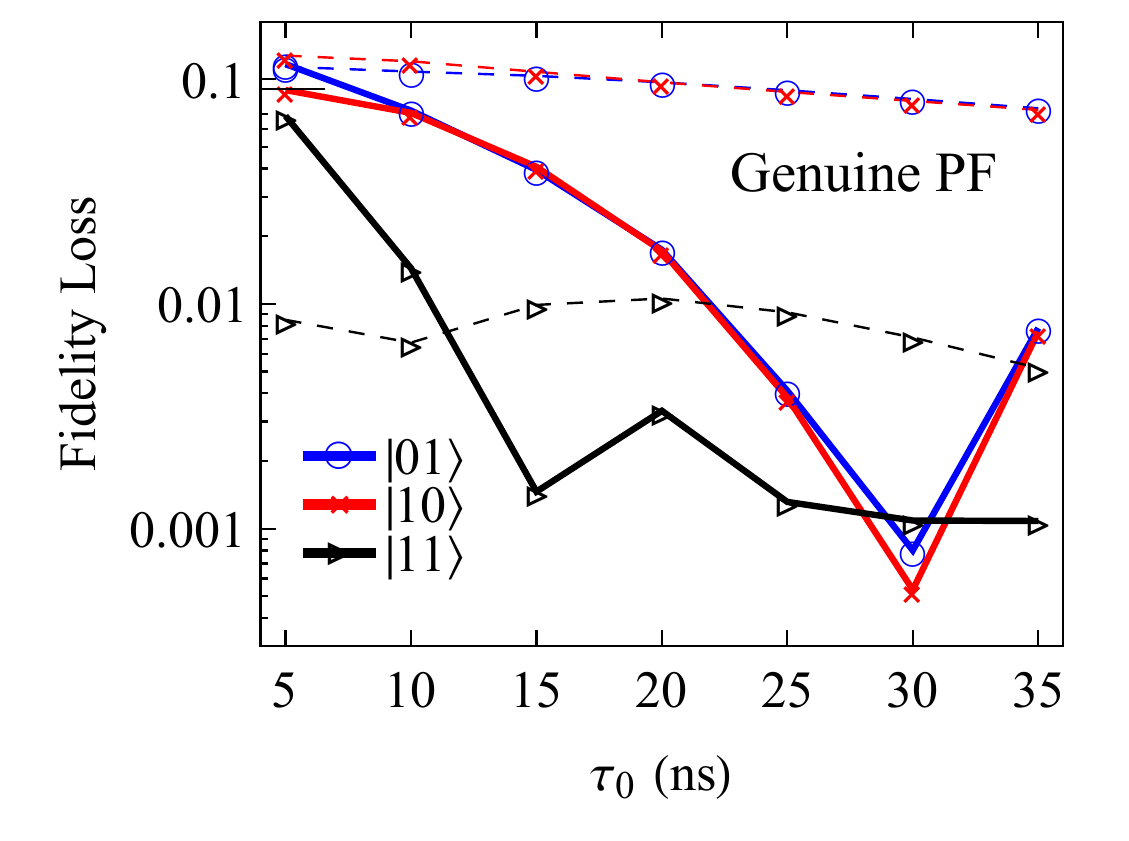}\put(-234,150){(d)}\\
	\includegraphics[width=0.41\textwidth]{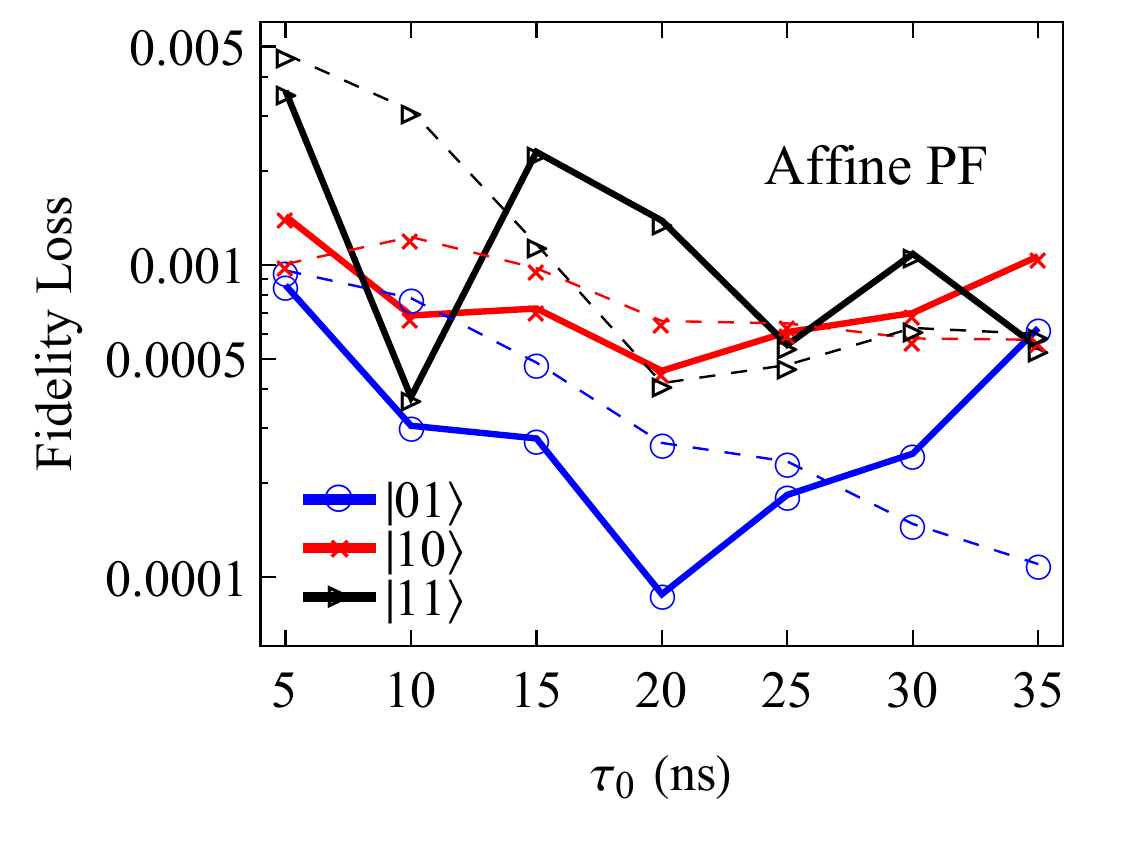}\put(-234,150){(e)}
	\vspace{-0.1in}
	\caption{(a) Sketches of two types of PF gate: start from idle mode, switch to the entangled mode and finally go back to the idle mode.
	(b) Coupler frequency switching protocols for genuine PF gate with the following two pulse envelopes: Hyperbolic tangent envelope (solid) and Flat-topped Gaussian envelope (dashed). (c) Coupler frequency switching protocols for affine PF gate with similar pulse envelopes.  Fidelity loss of the computational states for (d) genuine PF gate  and (e) affine PF gate due to leakage from the two pulse shapes. Coherence times of qubit/coupler/qubit are universal 200 $\mu$s. \label{fig:adia}}
\end{figure}

Figure~\ref{fig:adia}(c) shows similar two types of pulse envelope. The difference is that for affine PF gate idle coupler frequency $\wib$ is lower than $\omega_c^{\rm E}$. On-device 2 qubits are $ZZ$-free at $\wib/2\pi=$4.509~GHz. This regime with $1<|(\omega_{1,2}-\omega_c)/g|<10$ and static $ZZ$ freedom is the so-called Quasi-Dispersive Straddling regime and has been concluded as an optimal regime for fixed-frequency transmons~\cite{goerz2017charting}. To make a comparison, we tune the coupler frequency to make $\alpha_{ZX}$ also around 5~MHz but much closer to $\wib/2\pi$ --- at the frequency $\wE/2\pi$=4.530~GHz. 

Let us denote the total time it takes for the PF gate to start at  $\wi$ and return to it by  $t=2\tau_0+t_g$ in which $t_g$ is the microwave activation time in between two coupler frequency changes. Each coupler frequency change takes place during time $\tau_0$. By varying $\tau_0$, we evaluate fidelity loss during the switch between I and E modes: The circuit Hamiltonian is written in the qubit-coupler-qubit Hilbert space with a maximum of two excitations in the coupler by solving the master equation, we calculate the leakage both from qubits to the coupler and from computational subspace to higher levels. Finally, we determine the optimized pulse for frequency change. It is worth noting that here the coupler coherence time is comparable to qubits, and the fidelity loss is tiny. However, if the coupler coherence time is much shorter, one can see an obvious decrease in the fidelity of the PF gate; see Appendix~\ref{app:coco} for more details.

Figure~\ref{fig:adia}(d) shows the fidelity loss of states $|01\rangle$, $|10\rangle$, $|11\rangle$ for genuine PF gate by varying $\tau_0$ during the OFF-ON-OFF switch without an external drive. Both pulses show that the computational state fidelity increases by increasing $\tau_0$. However there is a difference between the two pulse performances.  The hyperbolic tangent envelope pulse which rapidly changes coupler frequency between I and E modes can further reduce the error by raising all computational state fidelities to above $99.9\%$ at  $\tau_0=30$~ns. For affine PF gate, the individual state fidelity loss is less than 0.1\% as shown in  Fig.~\ref{fig:adia}(e); in particular, the shortest rise/fall time for achieving 99.9\% total state fidelity is $\tau_0=10$~ns for the hyperbolic tangent envelope.

\vspace{-0.2in}
\subsection{Gate error during  external driving}
\vspace{-0.1in}

\begin{figure}[bp]
	\centering
	\includegraphics[width=0.48\textwidth]{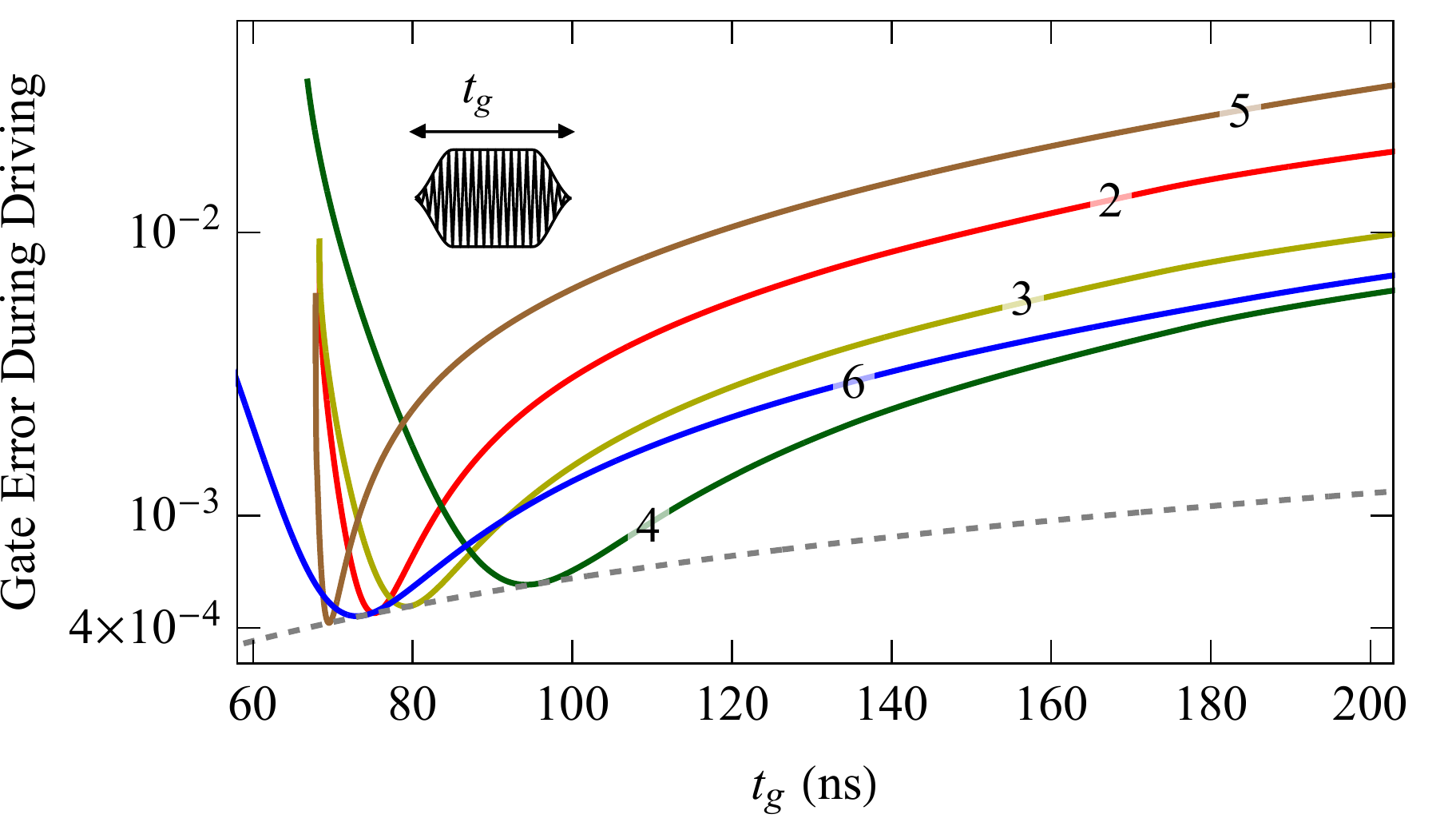}\put(-250,140){(a)}\put(-70,40){Genuine PF}\\
	\vspace{-0.1in}
	\includegraphics[width=0.48\textwidth]{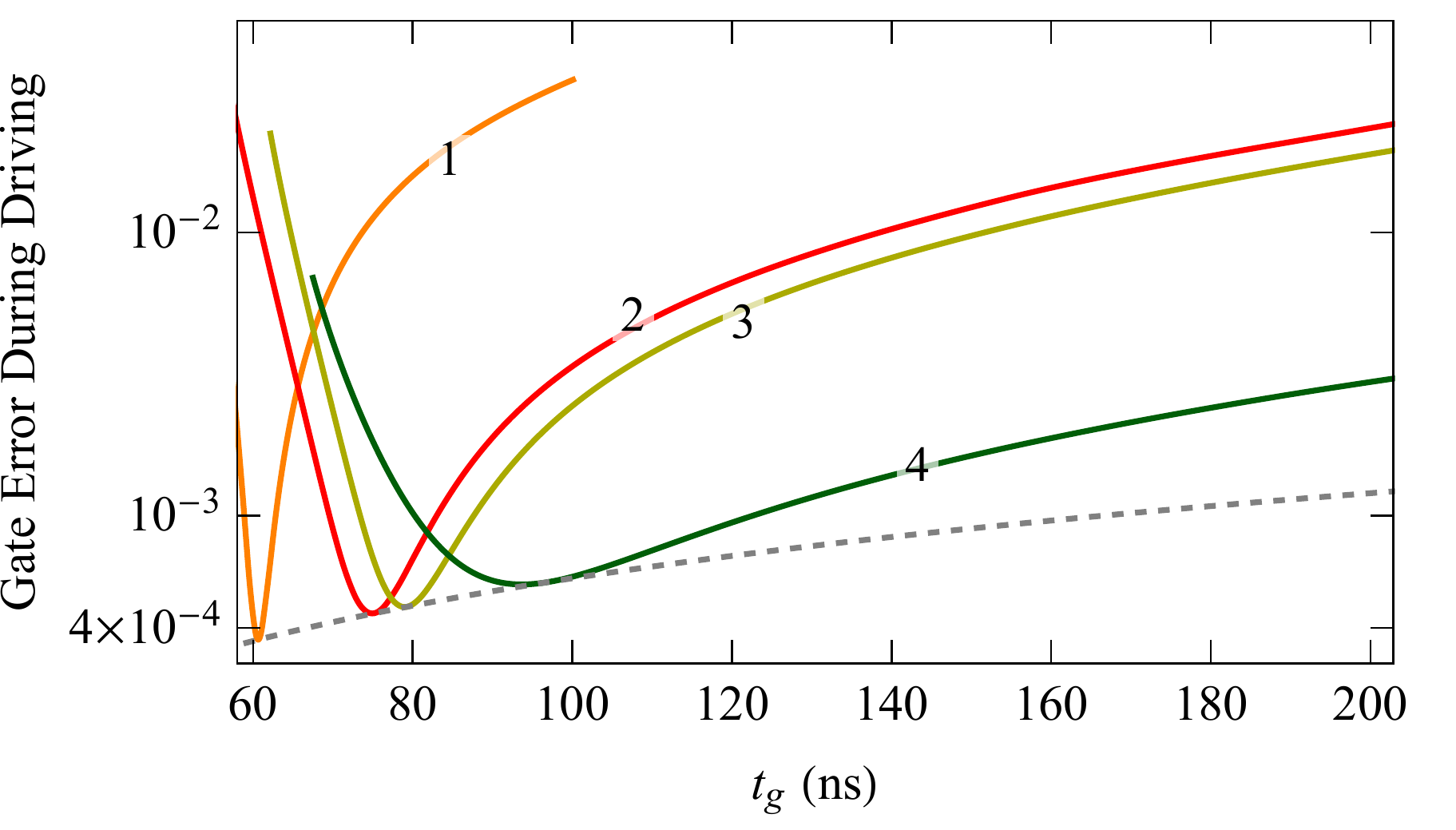}\put(-250,140){(b)}\put(-70,40){Affine PF}
	\vspace{-0.1in}
	\caption{(a) Genuine PF gate error during external driving versus gate length $t_g$. The coupler frequency is parked at $\omega_c/2\pi=4.8$ GHz. The inset plot is the round squared CR pulse shape with $\sim$20~ns rise and $\sim$20~ns fall times. (b) Affine PF gate error during external driving versus gate length $t_g$. The coupler frequency is parked at 4.472, 4.530, 4.658, and 4.731~GHz for devices 1-4. \label{fig:error}}
\end{figure}

In the two pulses discussed in Fig. \ref{fig:coupler2dy}(a) once the coupler frequency is changed to a lower value the circuit is ready to experience a $ZZ$-free $ZX$-interaction. This takes place by turning on the microwave during time $t_g$. The length of $t_g$ for $ZX$ gate is governed by microwave driving. During the time $t_g$ the qubit pairs enjoy absolute freedom from total $ZZ$ interaction. However, gate fidelity is limited by qubit coherence times. 

Let us indicate here that we do not consider the option of echoing the microwave driving as this doubles the gate length. Instead, we follow the recent practice at IBM where a single cross resonance driving is applied on qubits followed by a virtual $Z$ rotation~\cite{mckay2017efficient}. We also take the example of $ZX_{90}$ pulse for this typical analysis. In this pulse, the relation between $\alpha_{ZX}$ and the length of the flat top in the microwave pulse $\tau$ has been discussed before Section (\ref{sec:realization}), i.e. $\tau=1/4\alpha_{ZX}$. Here we consider the microwave pulses are round squared with $\sim$20~ns rise and $\sim$20~ns fall times as shown in the inset of Fig.~\ref{fig:error}. Thus the total $ZX$-gate length is $t_g=(40+1/4\alpha_{ZX}[\rm MHz])$ in nanoseconds. 

For the genuine PF gate, we consider the coupler frequency at E mode $\wE$ to be 4.8~GHz. With these parameters, the $ZX$-gate error rates are plotted for devices 2-6 in Fig.~\ref{fig:error}(a). All error rates have an expected behaviour as they show a minimum at a particular gate length $t_g$ where the error rate is as low as expected only from coherence time limitation where the device experiences total $ZZ$ freedom. Among the five devices 2-6 plotted, device 5 (a hybrid CSFQ-transmon) has the shortest gate length and after that stands device 6 (a pair of transmons with 200~MHz frequency detuning). Although the minimum error rate cannot be eliminated without perfecting individual qubits, it indicates the possibility of $ZX$ interaction gate with fidelity as high as $99.9\%$. Note that there is a limitation on the gate length behaving as a cutoff in Fig.~\ref{fig:error}, see Ref.~\cite{xu2021zz-freedom} for more details.

For the affine PF gate, we tune the coupler frequency such that the $ZX$ rate in devices 2-4 is the same as that in PF-G gate at corresponding freedom amplitude, and plot the error rate in  Fig.~\ref{fig:error}(b). The difference compared to the genuine PF gate is the coupler is only tuned in a narrow domain, e.g. 50 MHz, using a WTQ, which can effectively suppress decoherence from flux noise. Moreover, the required driving amplitude for the same gate duration is weaker, and the total $ZZ$ is smaller and less detrimental to the gate fidelity. Here we also study device 1 with the same qubit detuning as device 2. Figure~\ref{fig:error}(b) shows that device 1 enables the affine PF gate with a more substantial $ZX$ rate and therefore shorter gate length and less error rate.

Summing the error rate from both rise/fall times and decoherence, all together for device 2 one can estimate a minimum total time length of 95~ns long affine PF gate that takes Q1 and Q2 form a $ZZ$-free affine idle mode to entangled mode and returns it to original affine idle mode only by weakly tuning the coupler. The idle-to-idle error rate during the affine PF gate time of $t_g+2\tau_0$ will be about $0.1\%$. While for the genuine PF gate, the minimum gate length is 135 ns with 99.7\% gate fidelity.

The length of the PF gate can become even shorter if the microwave rise and fall times combine the two switching coupler frequency times. This can save up to 40~ns from the gate length for the example discussed above. Theoretically, such a time-saving needs careful analysis in optimal control theory, which goes beyond the scope of this paper; however experimentally it can be investigated.

\vspace{-0.2in}
\section{Summary}
\vspace{-0.1in}
To summarize, we propose an error-protected CR gate by combining idling and entangling gates. This gate can safely switch qubits states between idle and $ZX$-entangled modes, and once at both modes, quantum states do not accumulate conditional $ZZ$ phase error in time.  By zeroing $ZZ$ in both modes, qubits before, during, and after entanglement are safely phase-locked to their state. This gate can be realized in superconducting circuits by combining tunable circuit parameters and external driving in two ways: 1) At genuine idle mode, tuning circuit parameter makes qubits decoupled and therefore, the static $ZZ$ interaction vanishes. At entangled mode, the static $ZZ$ interaction is cancelled by a microwave-assisted $ZZ$ component so that qubits are left only with the operation of $ZX$-interaction; 2) At affine idle mode qubits are strongly coupled, but the level repulsions from both sides of computational space cancel each other. At entangled mode, qubits $ZX$-interact with zero-$ZZ$.

We evaluate a typical time length for the PF gate once its fidelity is only limited by qubit coherence times. In a complete operational cycle from OFF to OFF switch, passing once through an entangled mode, the affine PF gate is as short as 95~ns, with the overall error budget rate being about 0.1$\%$. 

We show that this gate is universally applicable for all types of superconducting qubits, such as all transmon or hybrid circuits,  and certainly not limited to frequencies in the dispersive regime. The PF gate will pave a new way to implement high-quality quantum computation in large-scale scalable quantum processors. 
\vspace{-0.2in}
\section*{Acknowledgement}
\vspace{-0.1in}
The authors thank Britton Plourde, Thomas Ohki, Guilhem Ribeill, Jaseung Ku, and Luke Govia for insightful discussions. We gratefully acknowledge funding from the German Federal Ministry 
of Education and Research within the funding program "Quantum Technologies - From Basic Research to the Market" (project GeQCoS), contract number
13N15685.

\appendix

\vspace{-0.2in}
\section{Comparing numerical and perturbation methods}\label{app compare}
\vspace{-0.1in}
The circuit Hamiltonian in the lab frame is written in the form of multilevel systems as 
\begin{eqnarray}\label{eq.hcoupler}
H_0=&&\sum_{i=1,2,c}\sum_n\omega_i(n_i)|n_i+1\rangle\langle n_i+1|+\sum_{i<j} \sum_n\nonumber\\
&&\sqrt{(n_i+1)(n_j+1)} g_{ij}\left(|n_i,n_j\rangle\langle n_{i}+1,n_{j}+1|\right.\nonumber\\
&&\left.-|n_{i}+1,n_j\rangle\langle n_i,n_{j}+1|+{ H.c.}\right),
\end{eqnarray}
where $\omega_i(n_i)=E_i(n_i+1)-E_i(n_i)$ and $\delta_i(n_i)=E_i(n_i+2)-2E_i(n_i+1)+E_i({n_i})$ with $E_i(n_i)$ being the bare energy of level $n$ for subsystem $i~(i=1,2,c)$. Especially, frequency and anharmonicity can be simplified as $\omega_i(0)=\omega_i$ and $\delta_i(0)=\delta_i$. 

\begin{figure}[h!]
	\centering
	\includegraphics[width=0.48\textwidth]{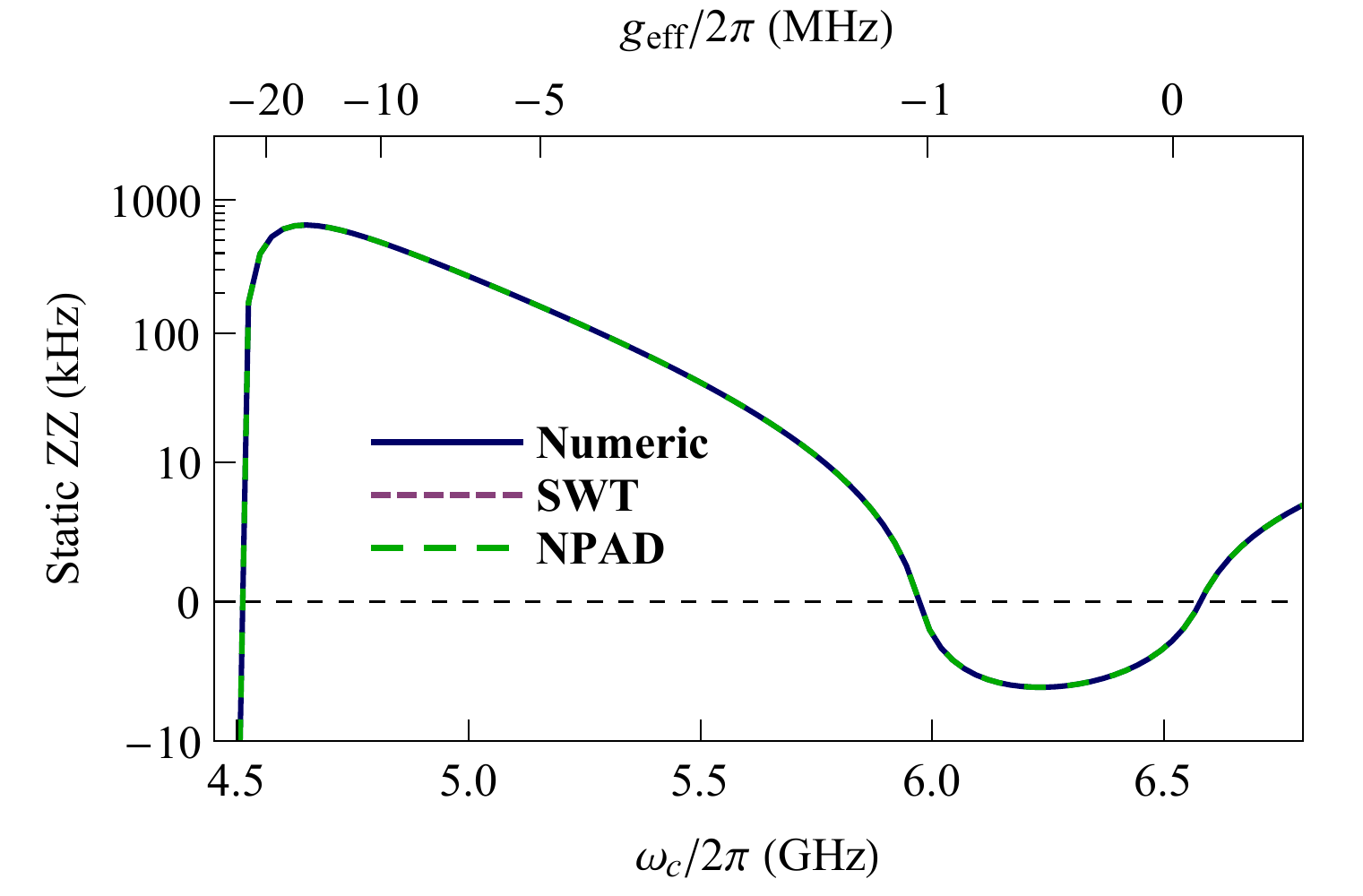}
  	\vspace{-0.3in}
	\caption{Static $ZZ$ interaction versus coupler frequency, SWT result is from Eq.~(\ref{eq.zeta}), NPAD makes use of the Jacobi iteration~\cite{li2021non-perturbative} and exact result is obtained by diagonalizing the Hamiltonian in Eq.~(\ref{eq.hcoupler}). The top axis is the corresponding effective coupling $g_{\rm eff}$. The used circuit parameters are the same as device 2 with $\alpha_1=\alpha_2=0.022$.}
	\label{fig:staticzz}
\end{figure}

 We evaluate static $ZZ$ on the seven devices listed in Table~\ref{tab:device} by fully diagonalizing the circuit Hamiltonian and plotting results in Fig.~\ref{fig:coupler2zz}. We also compare static $ZZ$ interaction in device 2 with the following three methods: numeric simulation (Numeric), NPAD~\cite{li2021non-perturbative} and SWT. Figure~\ref{fig:staticzz} shows the static $ZZ$ at the lower $x$ axis for the coupler frequency, as well as  $g_{\rm eff}$ at the upper $x$ axis for the effective coupling strength.

\vspace{-0.2in}
\section{Driven Hamiltonian}\label{app.d26} 
\vspace{-0.1in}
When microwave drive is on, CR driving Hamiltonian $H_d=\Omega\cos(\tilde{\omega}_2 t)\left(|n_1\rangle\langle n_{1}+1|+|n_1+1\rangle\langle n_1|\right)$ needs to be transferred to the same regime as the qubit Hamiltonian. In the rotating frame the total Hamiltonian is 
\begin{equation}
H_r=W^{\dagger}(\tilde{H_0}+\tilde{H_{ d}}) W-i W^{\dagger}\dot{W}.
\label{eq.Hr}
\end{equation}
where $\tilde{H_0}=U^{\dagger}H_0U$ with $U$ being the unitary operator that fully diagonalizes $H_0$, $\tilde{H_d}=U^{\dagger}H_dU$ and $W=\sum_{i=1,2,c}\sum_n \exp(-i\omega_d t \hat{n}_i)|n_i\rangle\langle n_i|$. 
For simplicity we assume $g_{1c}=g_{2c}=g$, $g_{12}=0$ and $\delta_1=\delta_2=\delta$, the transition rates in Eq.~(\ref{eq. Horig}) then are derived from perturbation theory and listed in Table~\ref{tab:lambda}.

In the entangled mode only qubits are encoded; we can further simplify the total Hamiltonian by decoupling the tunable coupler,  block diagonalizing the Hamiltonian and then rewriting it in terms of Pauli matrices as discussed in Sec.~\ref{gmode}. 

\begin{longtable}{@{\extracolsep{\fill}}cc@{}}
	\caption{Transition rates}
	\label{tab:lambda}
	\endfirsthead
	\endhead
	\hline\hline\\[-2.5ex]
	\centering
	$\lambda_1$&$ -g^2 \delta  /  2\Delta_{12} \Delta_{2} (\Delta_{12} +\delta)$\\
	$\lambda_2$&$ -g  \delta  / 2\Delta_{1} (\Delta_{1}+\delta)$\\
	$\lambda_3$&$ -\sqrt{2} g^2  \delta  /\Delta_{12} (\Delta_{12}-\delta) (\Delta_{2}+\delta)$\\
	$\lambda_4$&$ -g   /2 \Delta_{1} $\\
	$\lambda_5$&$ -g^2  \delta  /\sqrt{2} \Delta_{12} \Delta_{2} (\Delta_{12}+\delta)$\\ 
	$\lambda_6$&$ g  \delta  / \sqrt{2} \Delta_{1} (\Delta_{1}+\delta)$\\
	$\lambda_7$&$ -g^2   (\Delta_{2}+\delta_c) /2 \Delta_{12} \Delta_{2} (\Delta_{2}-\delta_c)$\\ 
	$\lambda_8$&$-g   / \sqrt{2} (\Delta_{1}-\delta_c)$\\ 
	$\lambda_9$&$-g  \delta /\sqrt{2}\Delta_{1}(\Delta_{1}+\delta)$\\ 
	$\lambda_{10}$&$ g^2 /\Delta_{2}(\Delta_{12}+\delta)$\\ 
	$\lambda_{11}$&$-g^2  \delta/\Delta_{12} (\Delta_{12}-\delta)(\Delta_{2}+\delta)$\\ 
	\hline\hline
\end{longtable}

\vspace{-0.2in}
\section{Dynamic $ZZ$ freedom}\label{sec:dzzf}
\vspace{-0.1in}

\begin{figure}[h!]
	\centering
	\includegraphics[width=0.48\textwidth]{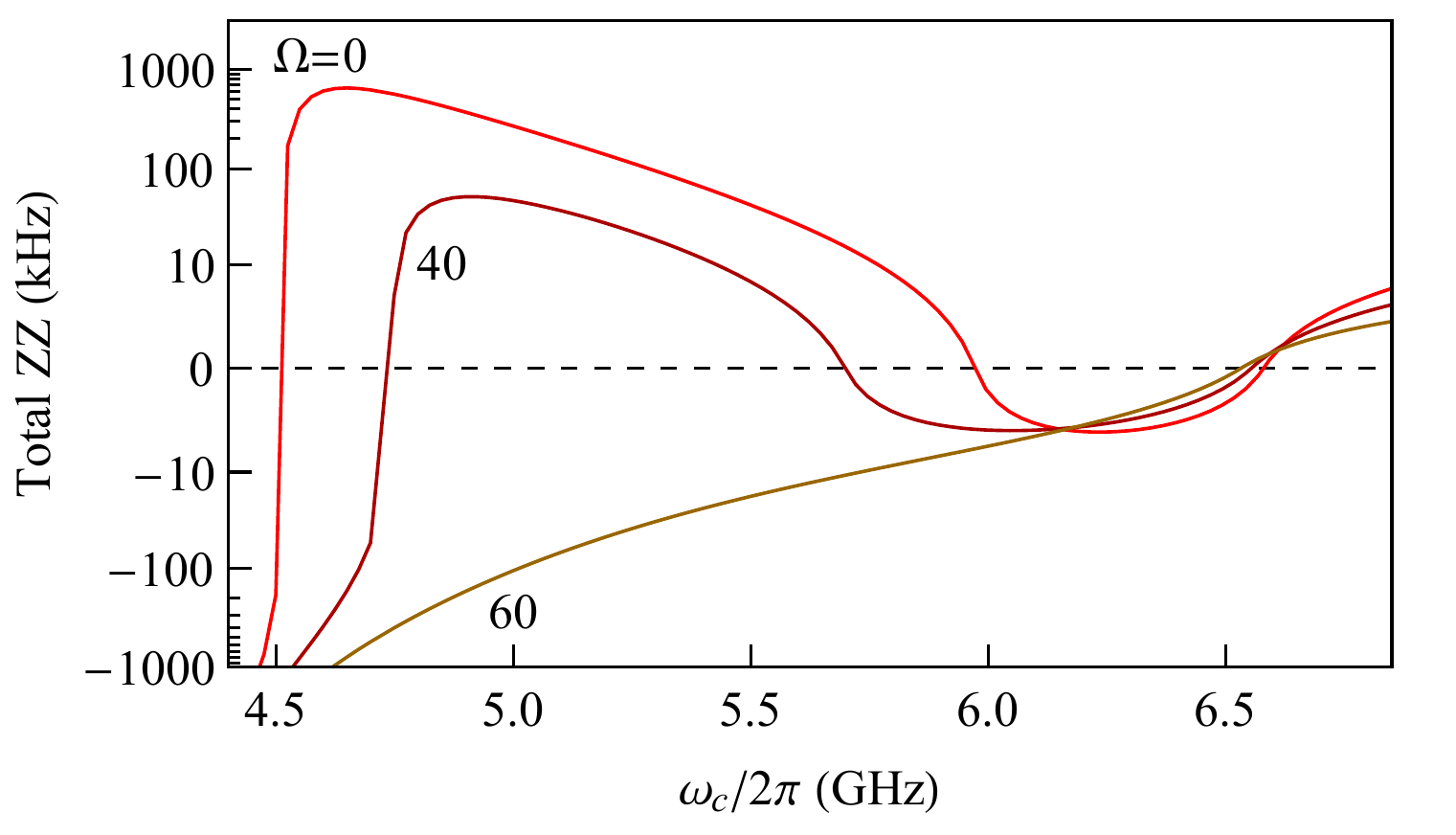}\put(-250,135){(a)}\put(-60,35){Device 2}\\
	\vspace{-0.05in}
	\includegraphics[width=0.48\textwidth]{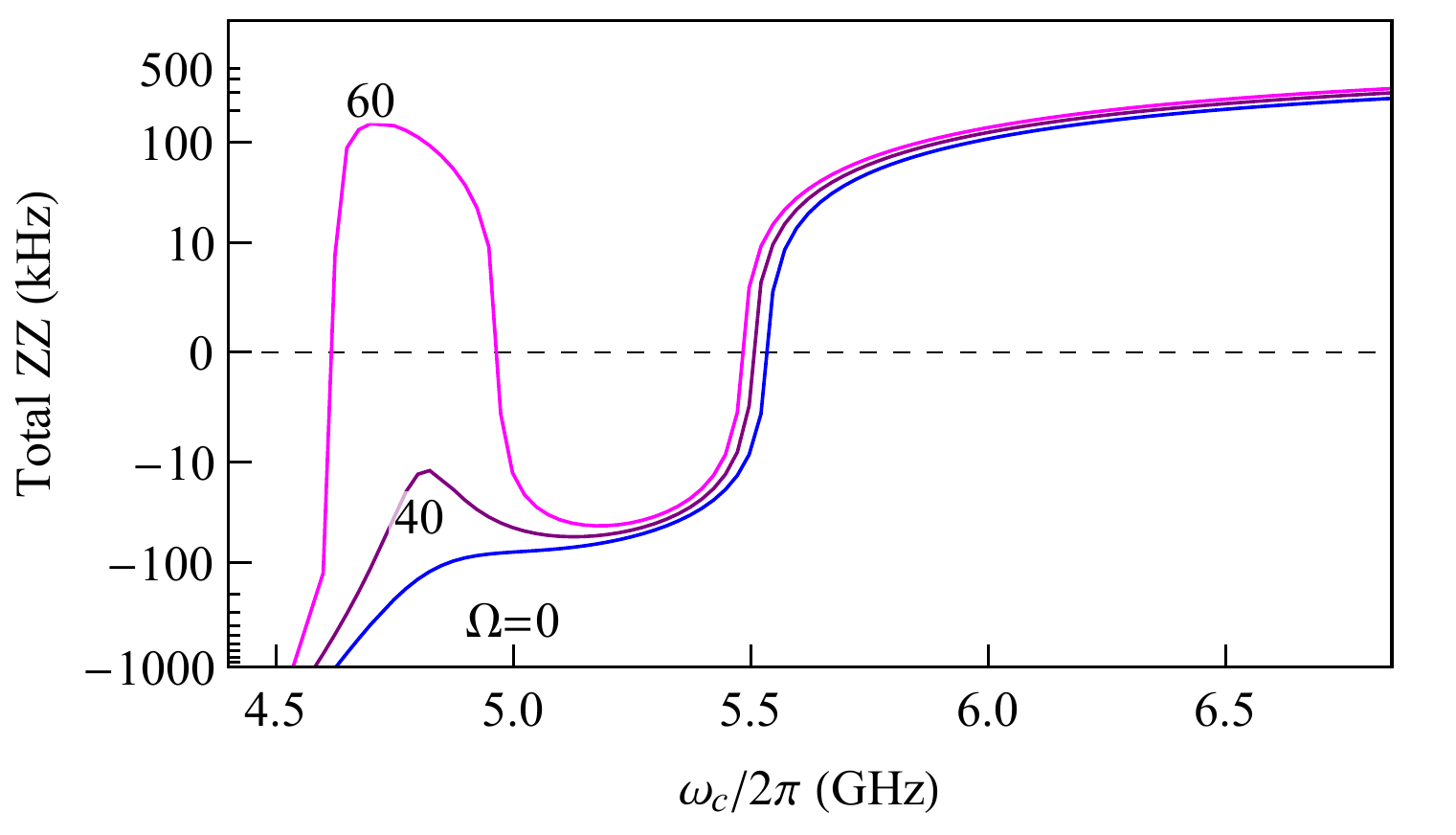}
	\put(-250,135){(b)}\put(-60,35){Device 6}
	\vspace{-0.2in}
	\caption{Total $ZZ$ interactions versus coupler frequency at different driving amplitude on (a) device 2 and (b) device 6.}
	\label{fig:dr26}
\end{figure}

Figure~\ref{fig:dr26} shows how the driving amplitude $\Omega$ impacts the total $ZZ$ interaction. In device 2, static $ZZ$ interaction exhibits three zero $ZZ$ points in terms of the coupler frequency. Increasing the driving amplitude makes the total $ZZ$ interaction smaller and annihilates two of the zero $ZZ$ points, leaving only one $\wia$. However, the behaviour of device 6 is the opposite; the external drive makes it possible to realize $ZZ$ freedom beyond the only $\wia$ point. 

\begin{figure*}
	\centering
	\includegraphics[width=0.5\textwidth]{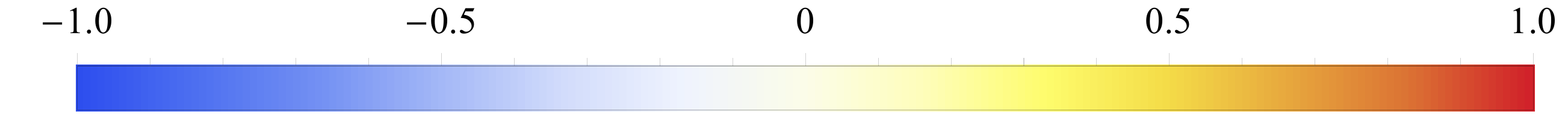}
	\put(-137,35){$\cos(\zeta \tau_p)$}\\
	\includegraphics[width=0.32\textwidth]{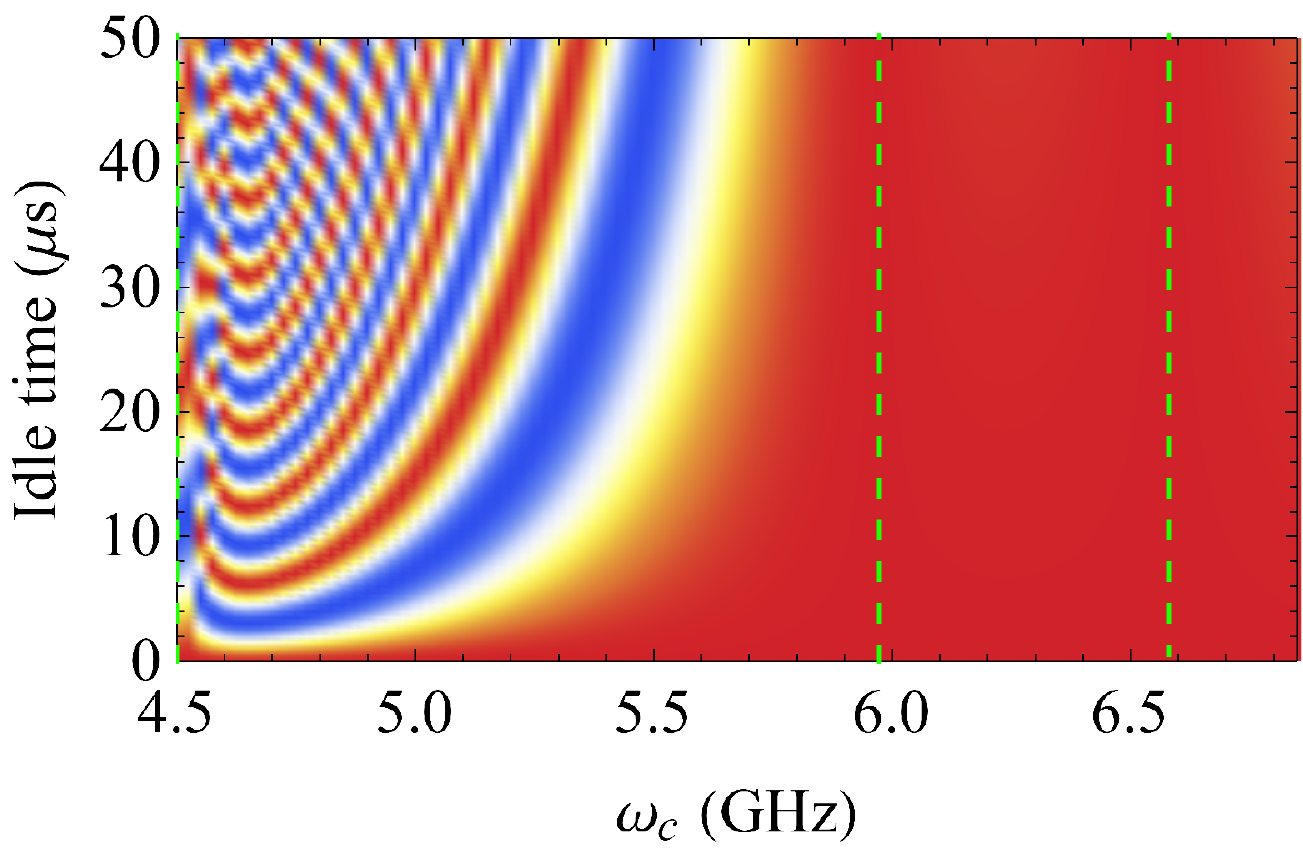}
	\includegraphics[width=0.32\textwidth]{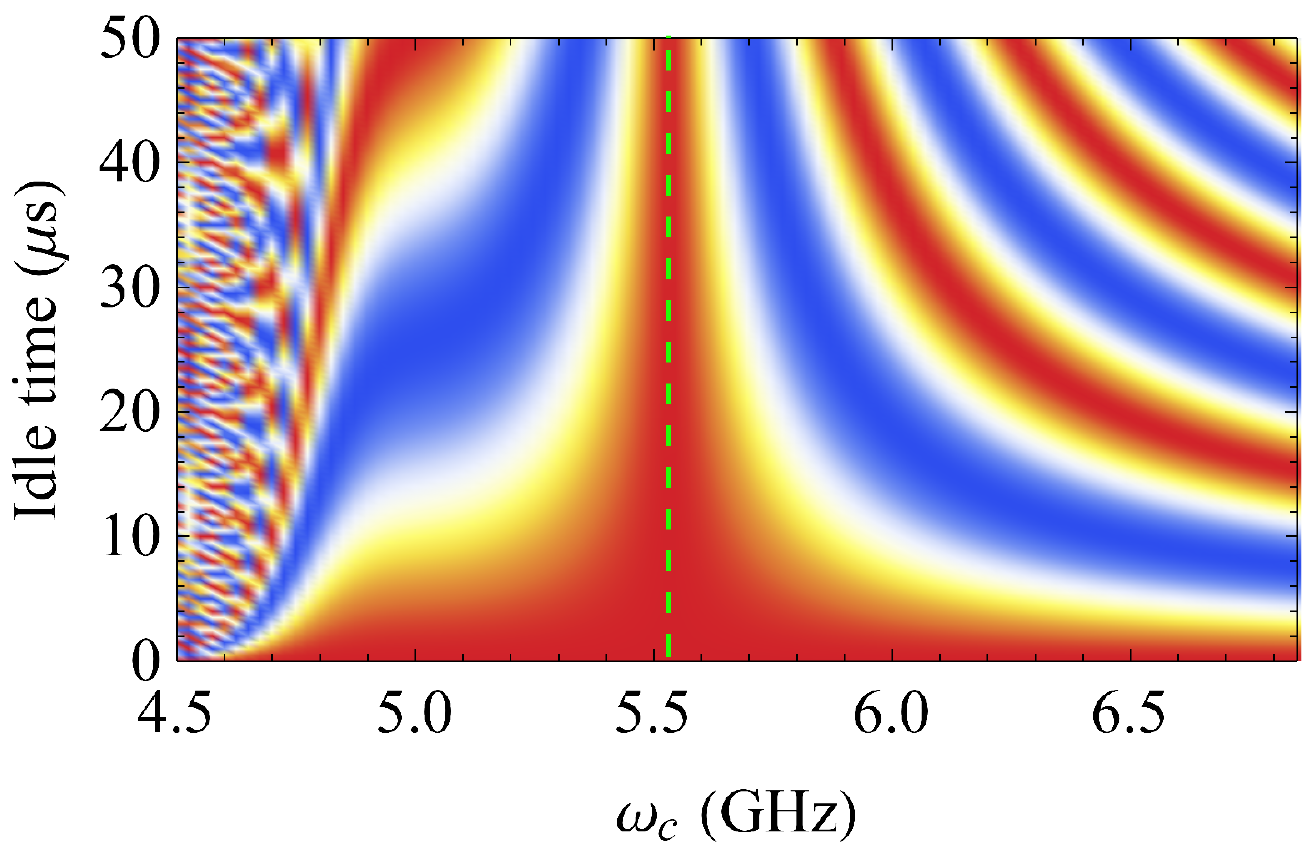}
	\put(-350,98){(a)}\put(-400,40){$\Omega=0$}\put(-245,35){\textcolor{green}{$g_{\rm eff}=0$}}
	\put(10,98){(b)}\put(20,40){$\Omega=0$}\put(-115,35){\textcolor{green}{$g_{\rm eff}=0$}}\vspace{-0.1in}\\
	\includegraphics[width=0.32\textwidth]{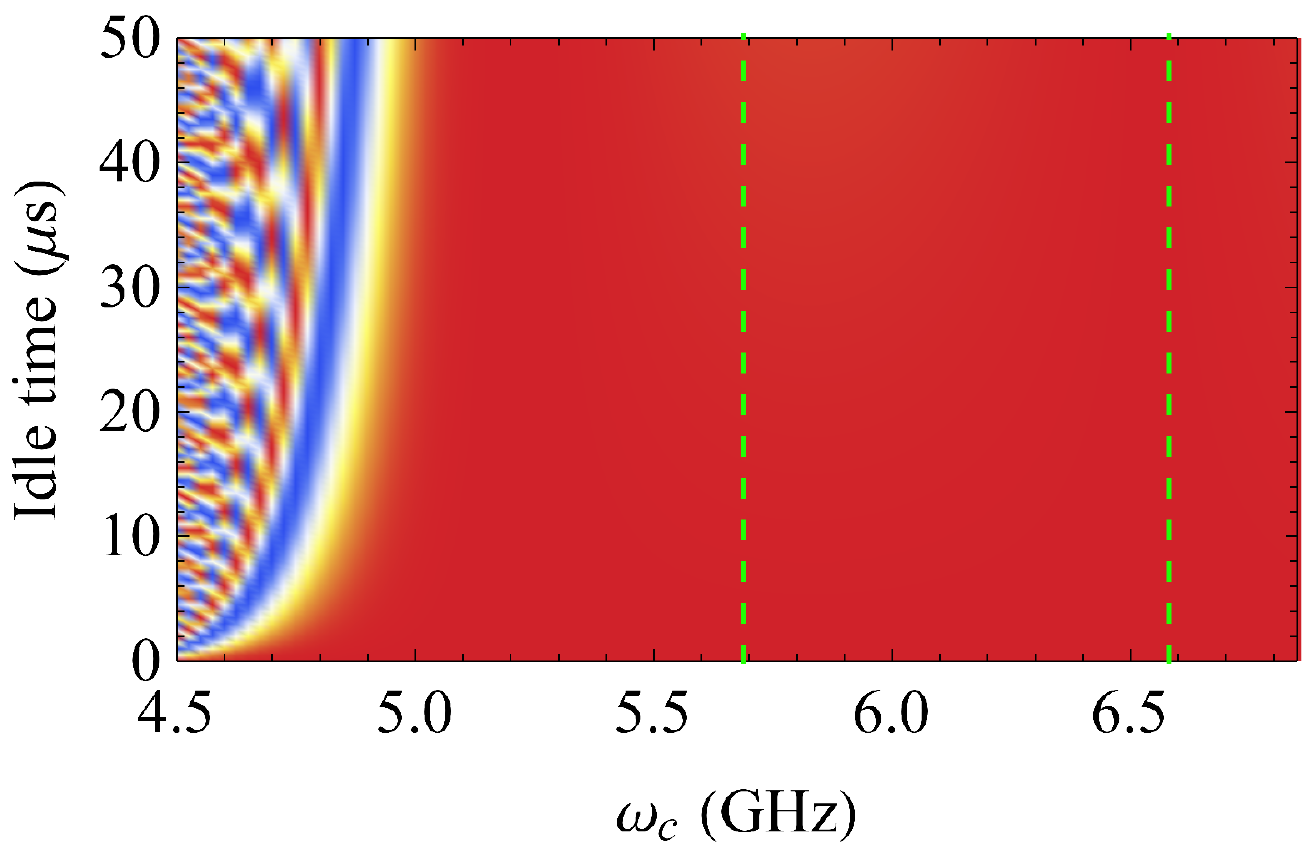}
	\includegraphics[width=0.32\textwidth]{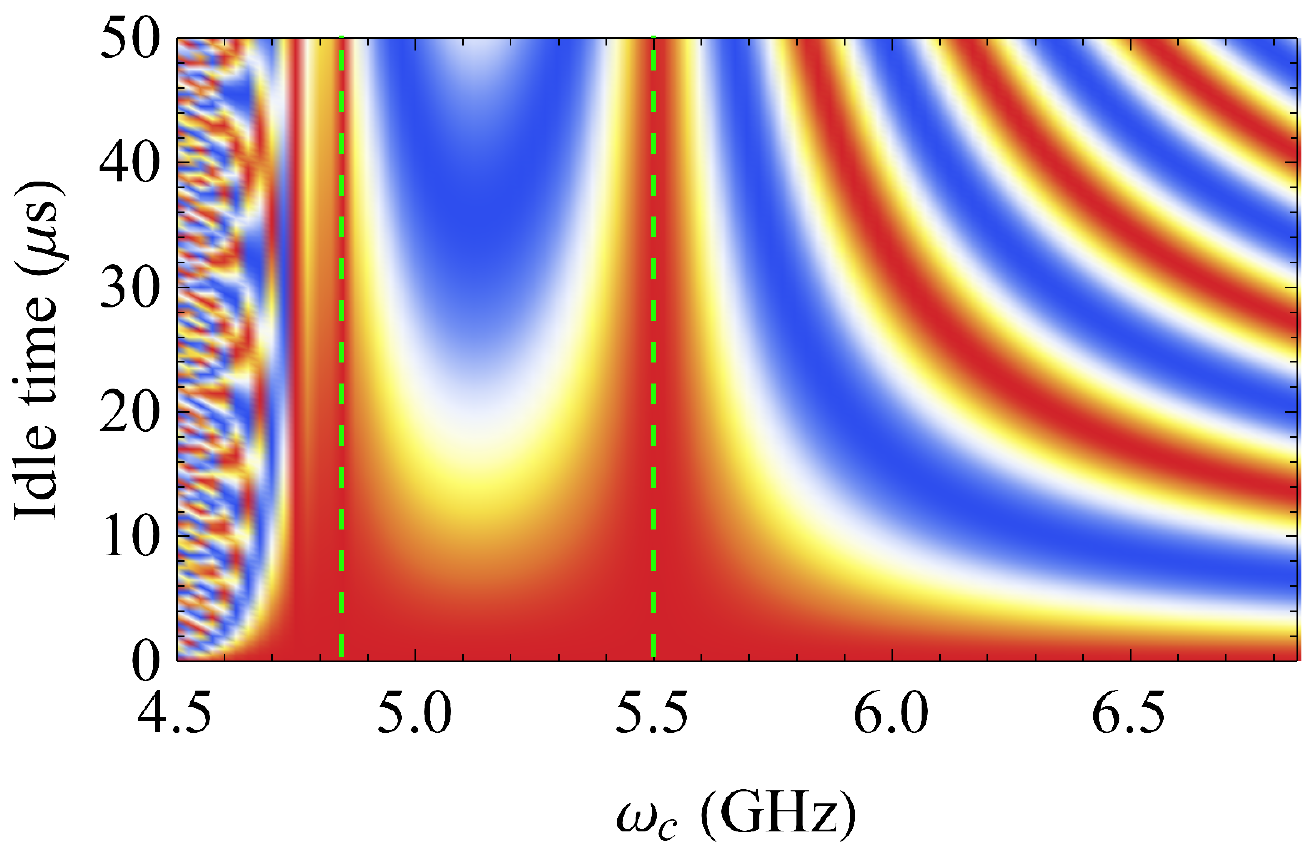}
	\put(-350,98){(c)}\put(-400,40){$\Omega=47$ MHz}
	\put(10,98){(d)}\put(20,40){$\Omega=42.3$ MHz}\vspace{-0.0in}\\
	\includegraphics[width=0.32\textwidth]{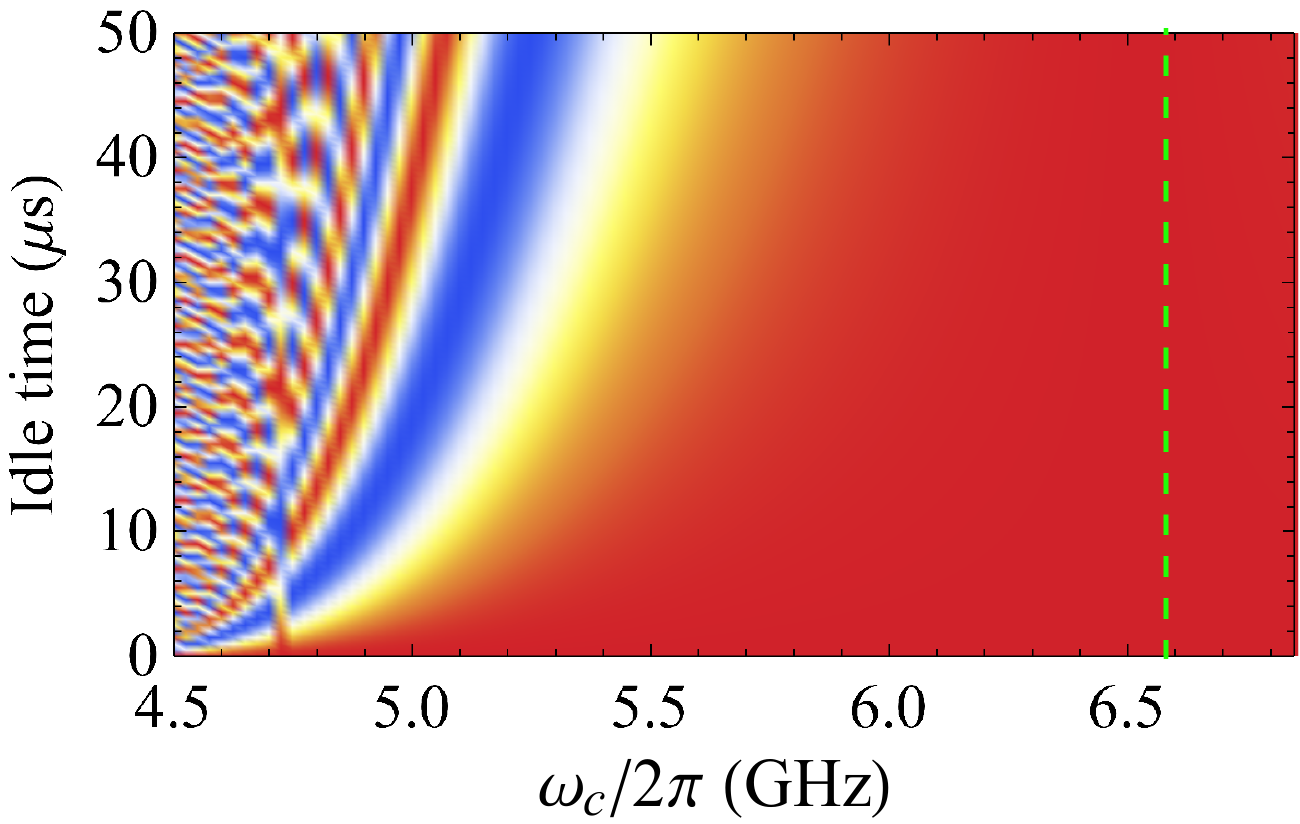}
	\includegraphics[width=0.32\textwidth]{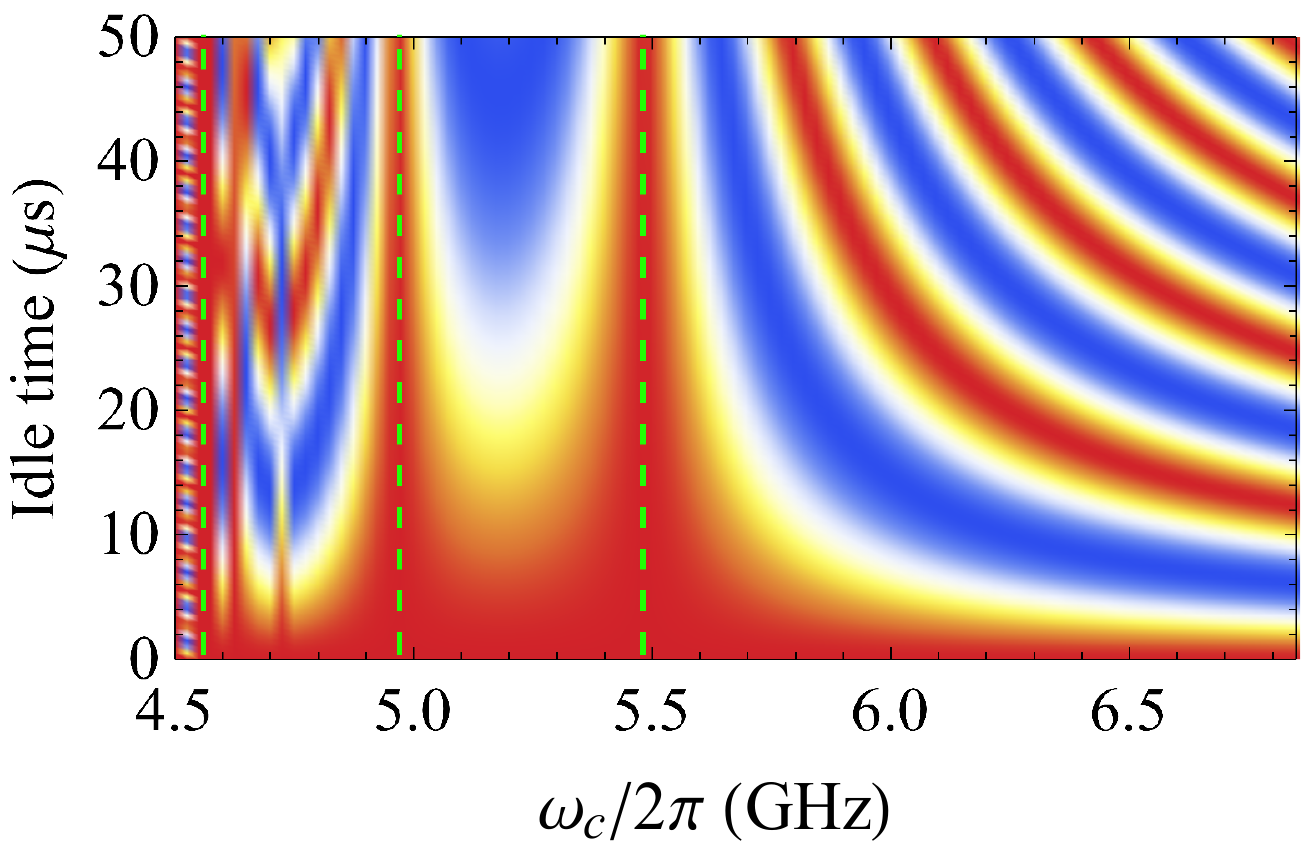}
	\put(-350,98){(e)}\put(-400,40){$\Omega=60$ MHz}
	\put(10,98){(f)}\put(20,40){$\Omega=60$ MHz}\\
	\vspace{-0.1in}
	\caption{Accumulated conditional phase error on computational states on (a,c,e) device 2 at driving amplitude $\Omega=0$, $\Omega=47$~MHz and $\Omega=60$~MHz, and on (b,d,f) device 6 at driving amplitude $\Omega=0$, $\Omega=42.3$~MHz and $\Omega=60$~MHz, respectively. Green dashed lines indicate the $ZZ$-free coupler frequency.}
	\label{fig:pop}
	\vspace{-0.2in}
\end{figure*}

To show how the computational states accumulate conditional phase error, we plot $\exp(i\zeta\tau_p)$ in devices 2 and 6 during the idling periods of duration $\tau_p$ in Fig.~\ref{fig:pop}. Usually such fringes can be measured by performing a Ramsey-like experiment on the target qubit to validate the $ZZ$ cancellation~\cite{ni2021scalable}. 

One can see that device 2 does not accumulate conditional phase at two additional coupler frequencies beyond $g_{\rm eff}=0$ as shown in Fig.~\ref{fig:pop}(a). Figure~\ref{fig:pop}(c) shows that these two $ZZ$-free coupler frequency points reduce to one at the critical amplitude $\Omega=47$ MHz. Above this amplitude, i.e. $\Omega=60$ MHz in Fig.~\ref{fig:pop}(e), the circuit can be free from parasitic $ZZ$ interaction only at $\wia$, indicating that the idle mode is robust against external driving. However, device 6 shows the opposite phenomenon. At the idle mode device 6 is only $ZZ$-free at  $\wia$ as shown in Fig.~\ref{fig:pop}(b). When driving amplitude is increased, $ZZ$-freedom can be found in additional coupler frequencies as shown in Fig.~\ref{fig:pop}(d) and \ref{fig:pop}(f).

\vspace{-0.2in}
\section{Impact of Higher order correction}\label{app.higherorder}
\begin{figure}[ht]
	\centering
	\includegraphics[width=0.24\textwidth]{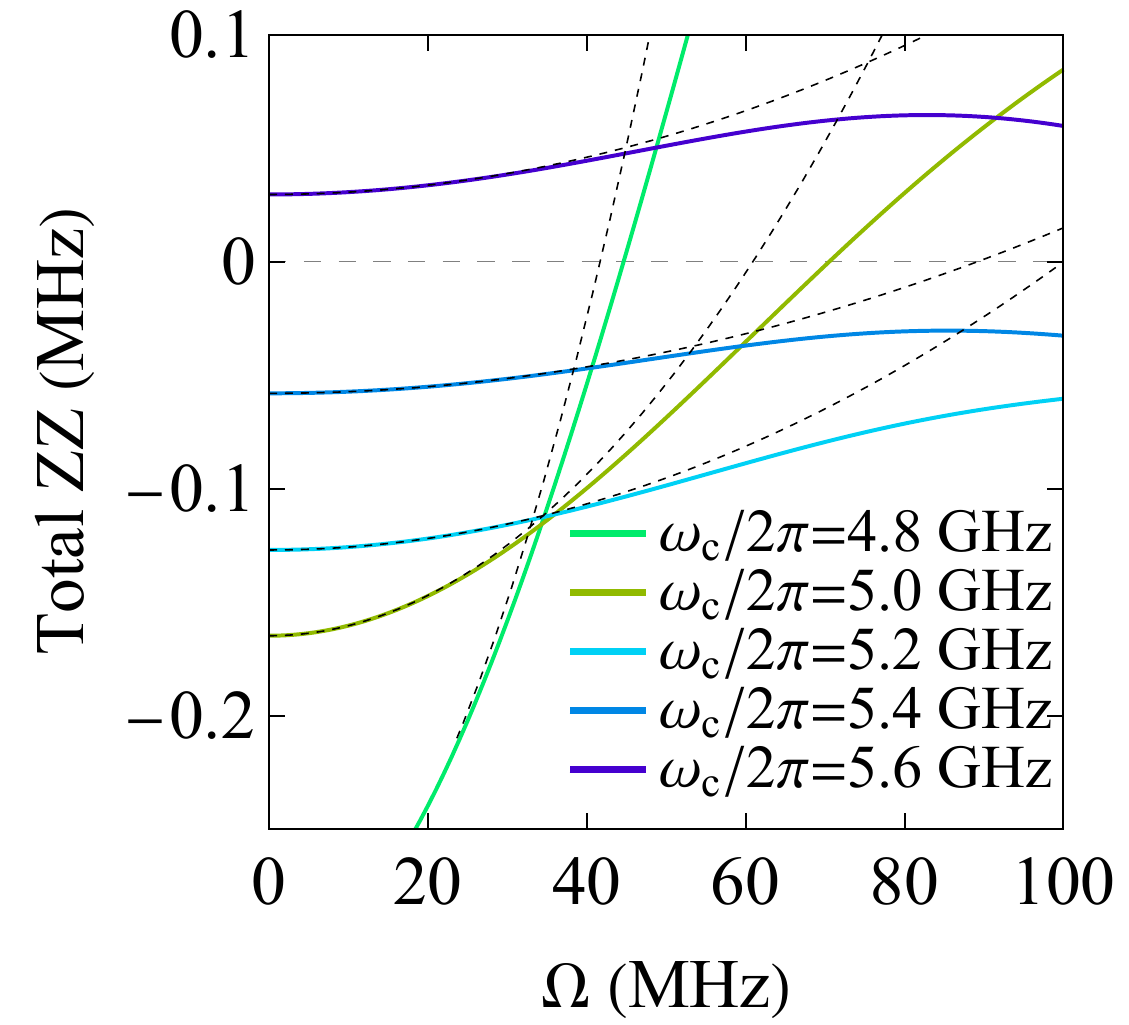}\hspace{-0.1in}
	\includegraphics[width=0.24\textwidth]{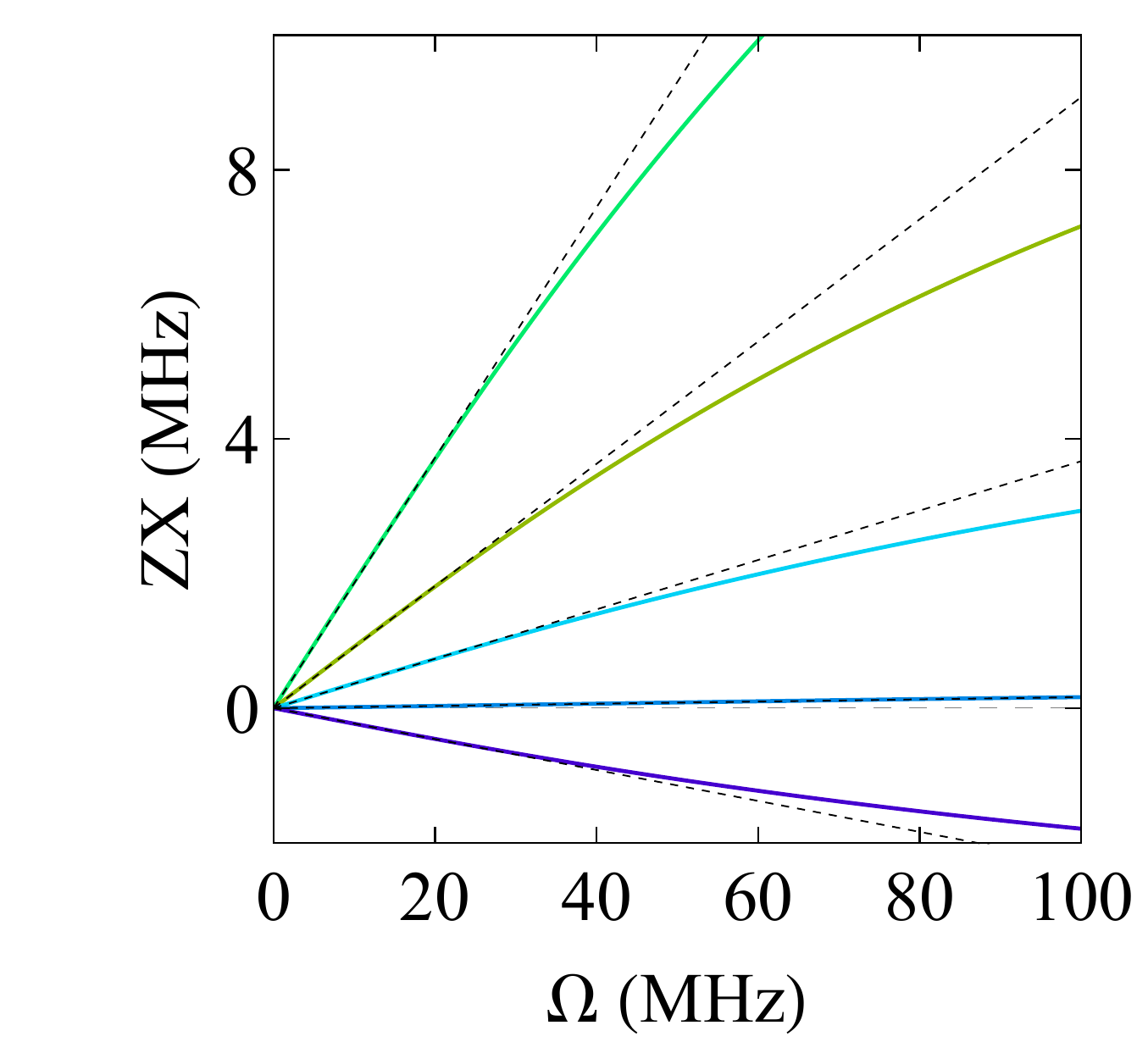}
	\put(-240,105){(a)}\put(-115,105){(b)}\\
	\includegraphics[width=0.24\textwidth]{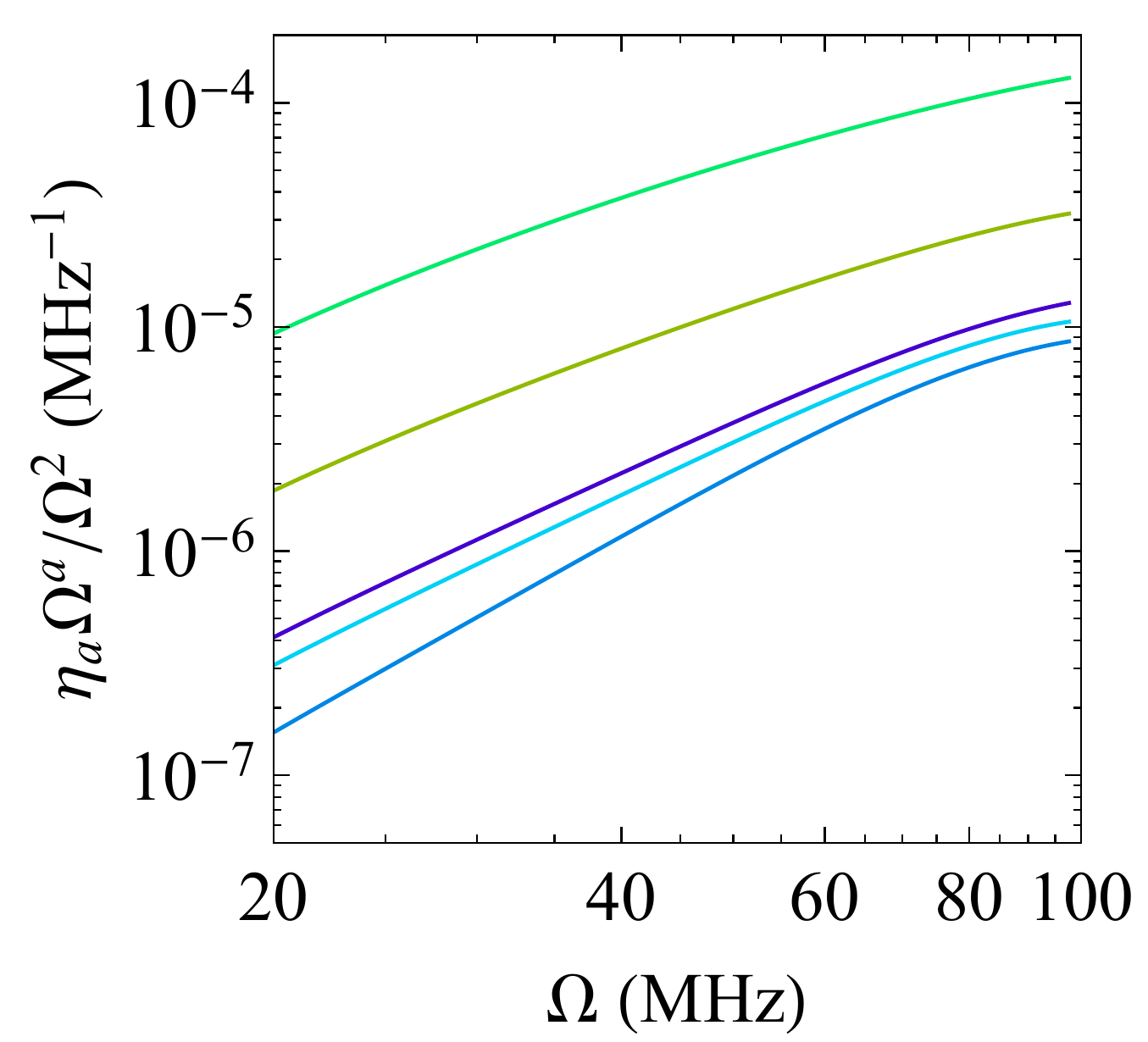}\hspace{-0.1in}
	\includegraphics[width=0.24\textwidth]{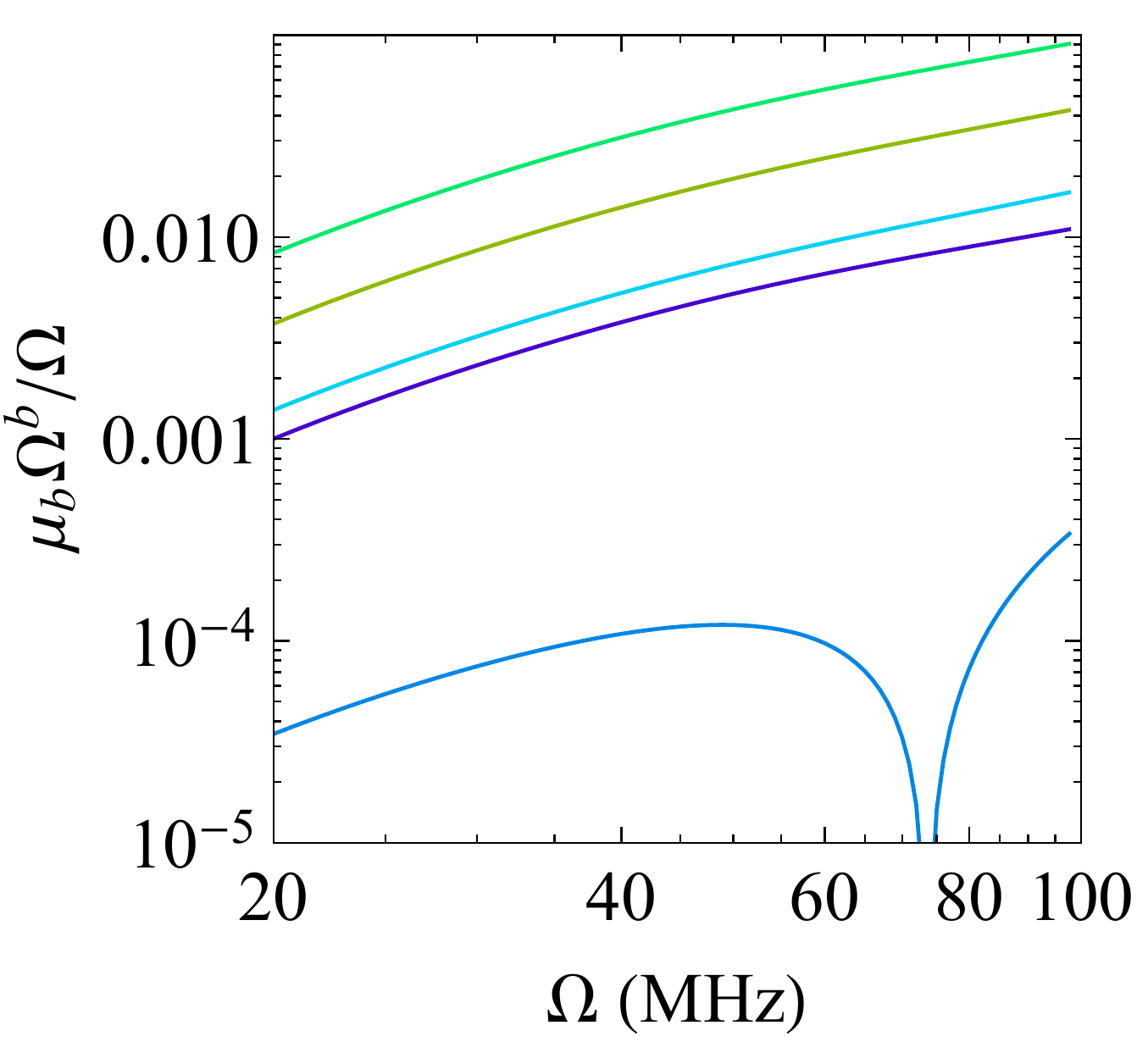}
	\put(-240,105){(c)}\put(-115,105){(d)}\\
	\vspace{-0.1in}
	\caption{In device 6 (a) Total $ZZ$ interaction versus driving amplitude. Dashed lines indicate the trend concerning only the sum of static and quadratic components (b) $ZX$ rate versus driving amplitude at different coupler frequencies, with dashed lines being the linear trend. (c) Higher order correction of $ZZ$ interaction versus driving amplitude on the log scale.  (d) Higher order correction of $ZX$ rate on the log scale.}
	\label{fig:hfactor}
\end{figure}

\vspace{-0.1in}
Figure~\ref{fig:hfactor}(a) and~\ref{fig:hfactor}(b) show total $ZZ$ interaction and $ZX$ rate in device 6 at different coupler frequencies. Dashed lines indicate the trend of the Pauli coefficients without higher order correction ($a,b=0$). However, in reality corrections from higher levels contribute such that $ZZ$ curves become flatter and finally purely negative with increasing coupling frequency. While the $ZX$ rate decreases from positive to negative continuously due to the sign change of $g_{\rm eff}$. The normalized higher-order terms are evaluated and plotted in Fig.~\ref{fig:hfactor}(c) and~\ref{fig:hfactor}(d). In the logarithmic scale, these higher-order terms are almost linear at lower driving amplitude and become flatter with increasing driving amplitude $\Omega$; the slopes also rise with the coupler frequency. Moreover, when the coupler frequency is tuned to be around $\wia$, effective coupling $g_{\rm eff}$ is relatively weak and will change its sign. Since $\eta_2\appropto g_{\rm eff}^2$ and $\mu_1\appropto g_{\rm eff}$ are extremely small in the vicinity of the idle coupler frequency $\omega_c^{\rm I}$, higher-order terms contribute dominantly at weak driving amplitude. 

\vspace{-0.2in}
\section{Quadratic factor}
\vspace{-0.1in}

\begin{figure}[ht]
	\centering
	\includegraphics[width=0.48\textwidth]{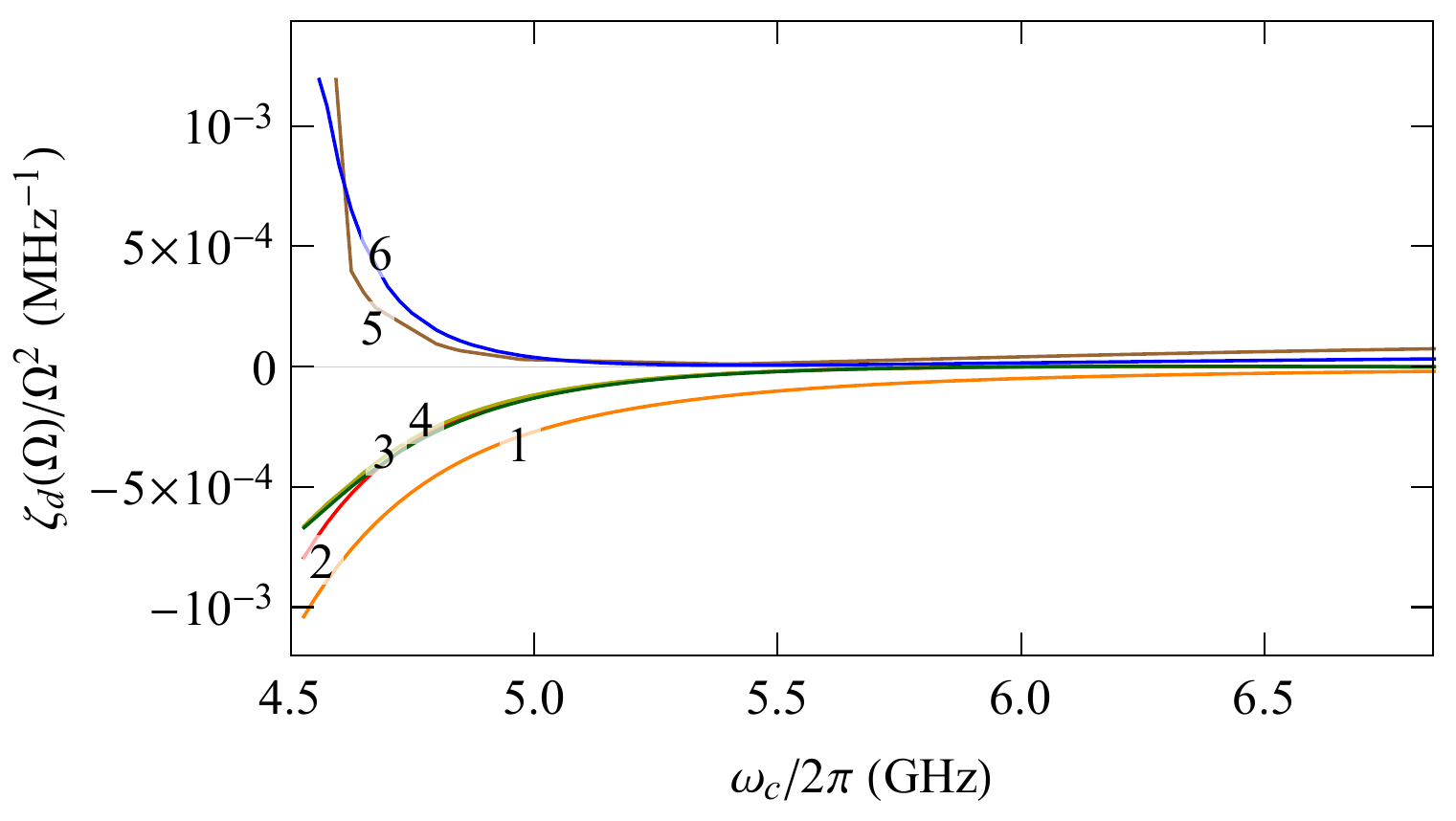}\put(-240,135){(a)}\\
	\vspace{-0.051in}
	\includegraphics[width=0.48\textwidth]{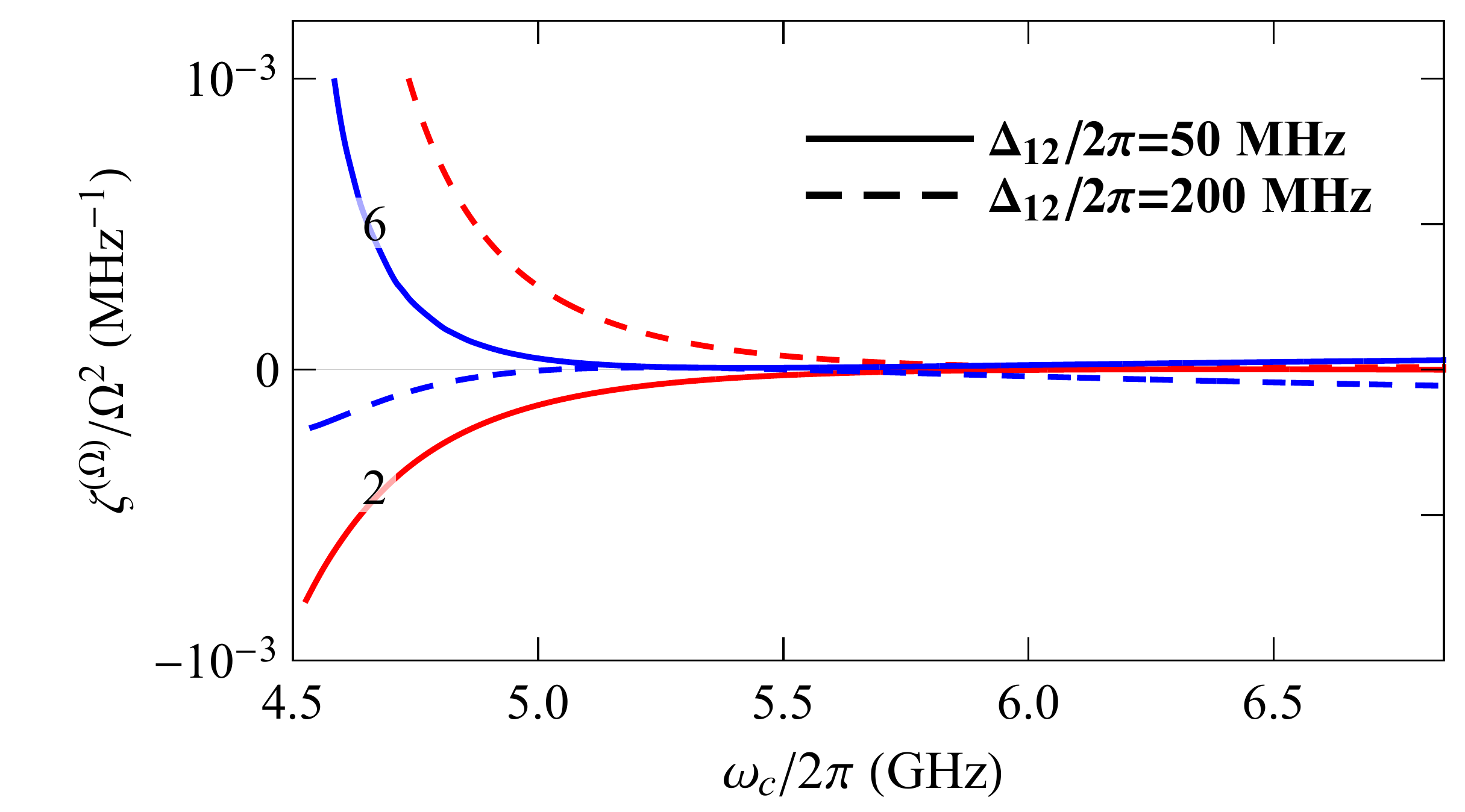}\put(-240,135){(b)}
	\vspace{-0.15in}
	\caption{(a) Quadratic factor $\eta$ as a function of coupler frequency in devices 1-6. (b) Quadratic factor $\eta$ as a function of coupler frequency in devices 2 and 6 at the qubit detuning $\Delta_{12}/2\pi=50$ MHz and $\Delta_{12}/2\pi=200$ MHz. }\label{fig:eta}
	\end{figure}

Naively, we can assume the dynamic $ZZ$ interaction is proportional to $\Omega^2$ as the coupler frequency is away from $\wia$. Figure~\ref{fig:eta}(a) shows that such normalized driven part $\zeta_d/\Omega^2$ in device 1-6 dramatically increases when the coupler frequency is close to the qubits, but approaches zero at higher $\omega_c$. In devices 1-4, the sign of quadratic factor is always negative, while in devices 5 and 6, it is positive. Figure~\ref{fig:eta}(b) shows that the sign of dynamic $ZZ$ interaction changes with the qubit-qubit detuning with respect to perturbatively $\zeta_d\appropto1/(\Delta_{12}+\delta_1/2)$.
	
\vspace{-0.2in}
\section{Impact of Coupler Decoherence}\label{app:coco}
\vspace{-0.1in}

\begin{figure}[b!]
	\centering
	\includegraphics[width=0.395\textwidth]{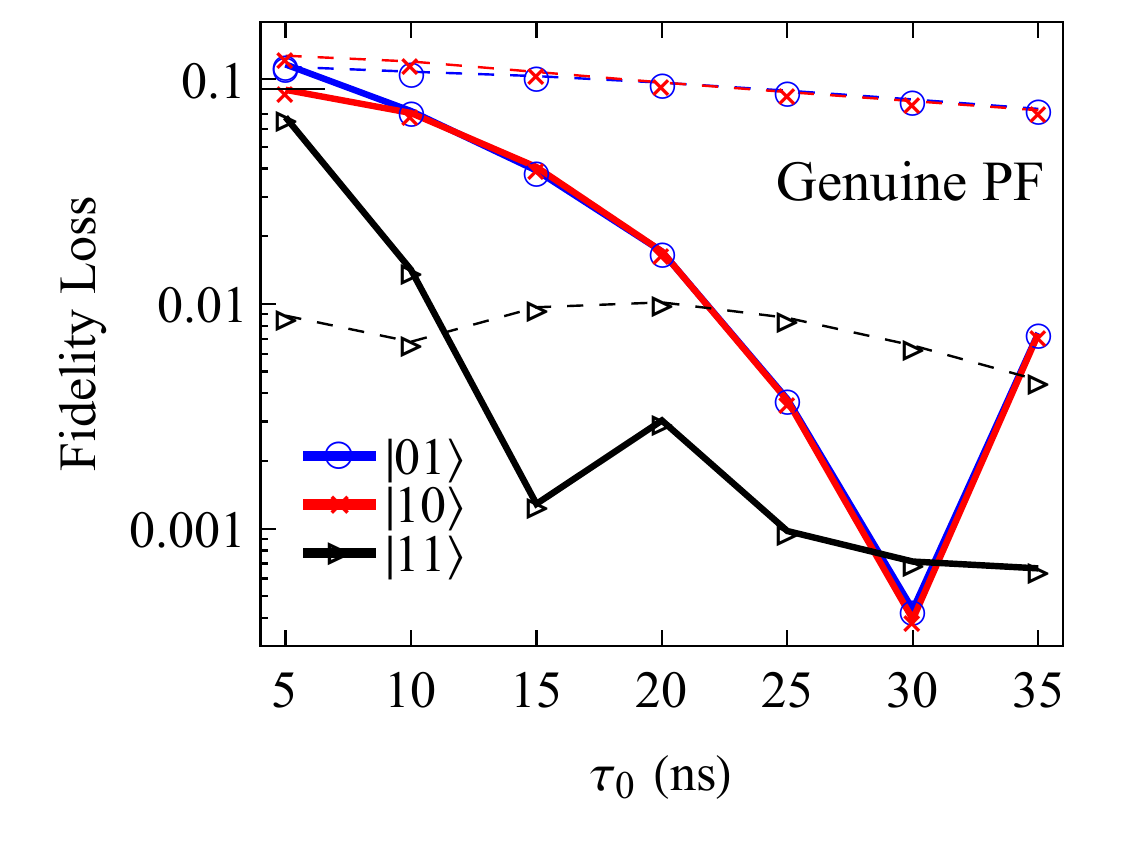}\put(-205,145){(a)}\\
	\vspace{-0.1in}
	\includegraphics[width=0.395\textwidth]{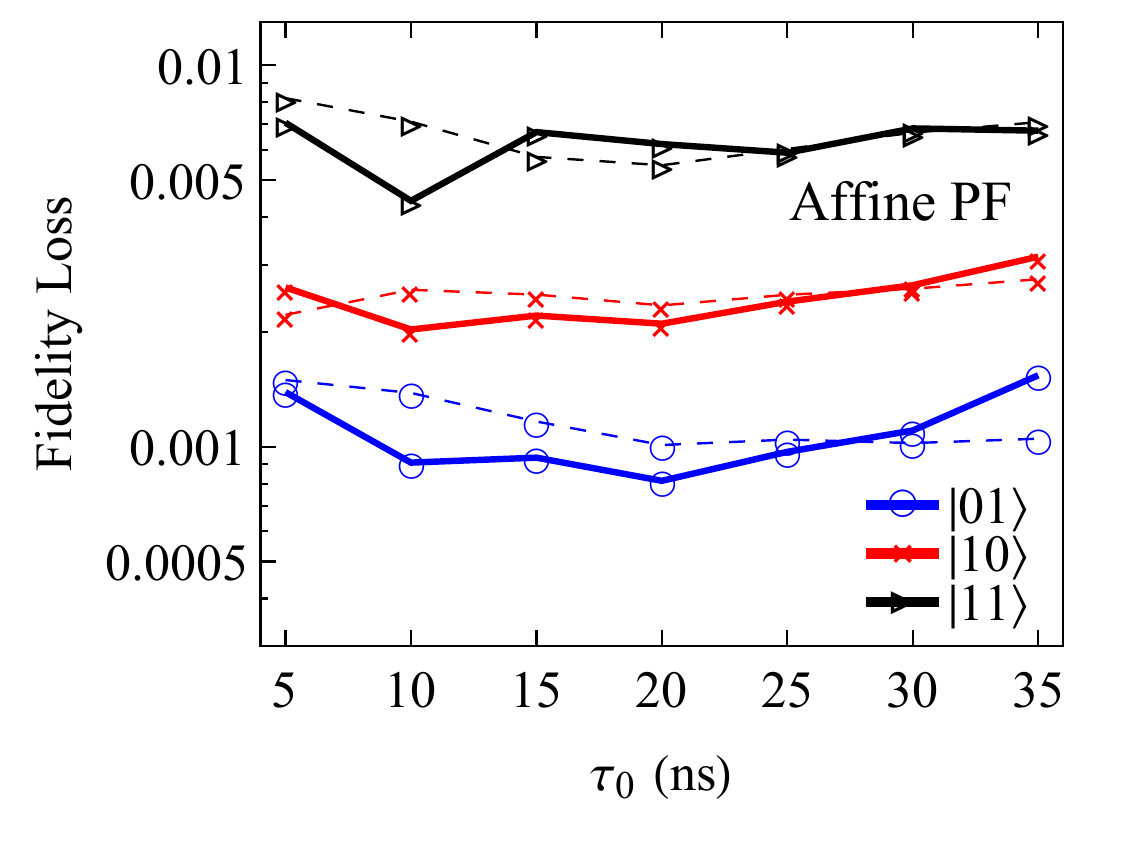}\put(-205,145){(b)}\\
		\vspace{-0.1in}
	\includegraphics[width=0.395\textwidth]{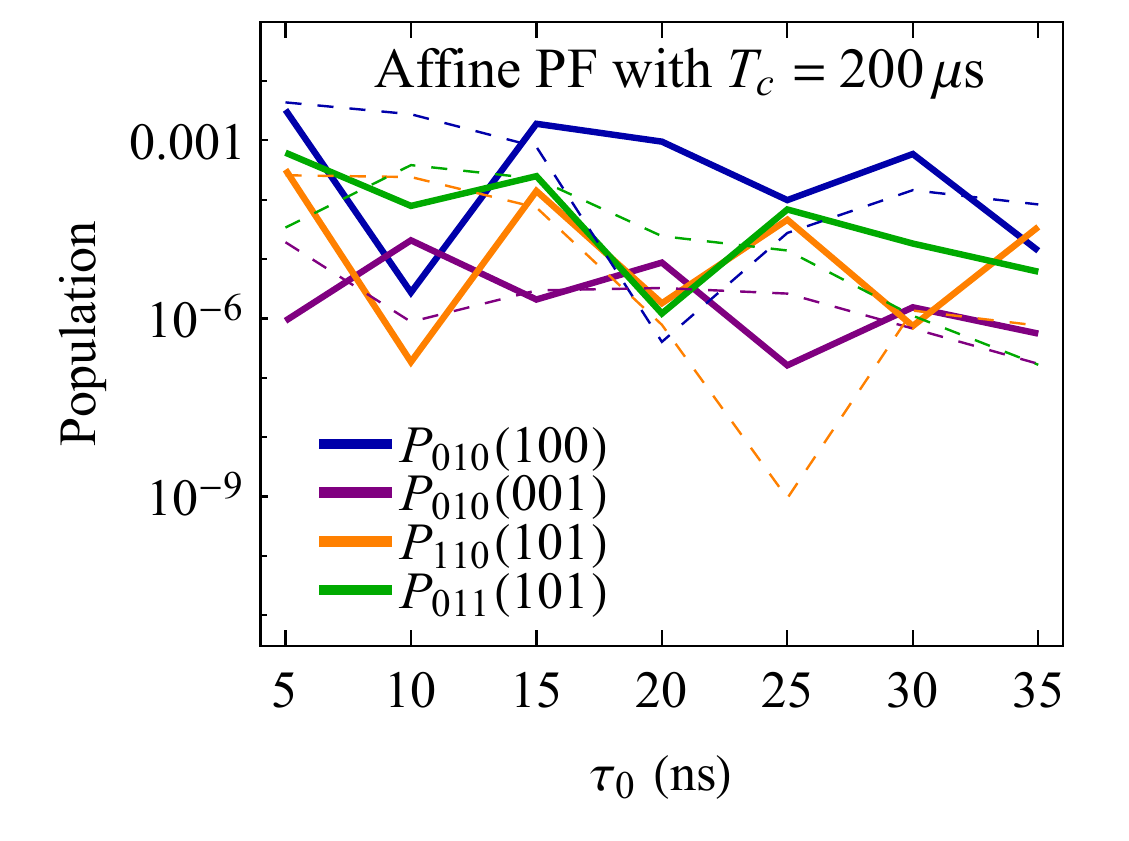}\put(-205,145){(c)}\\
		\vspace{-0.1in}
	\includegraphics[width=0.395\textwidth]{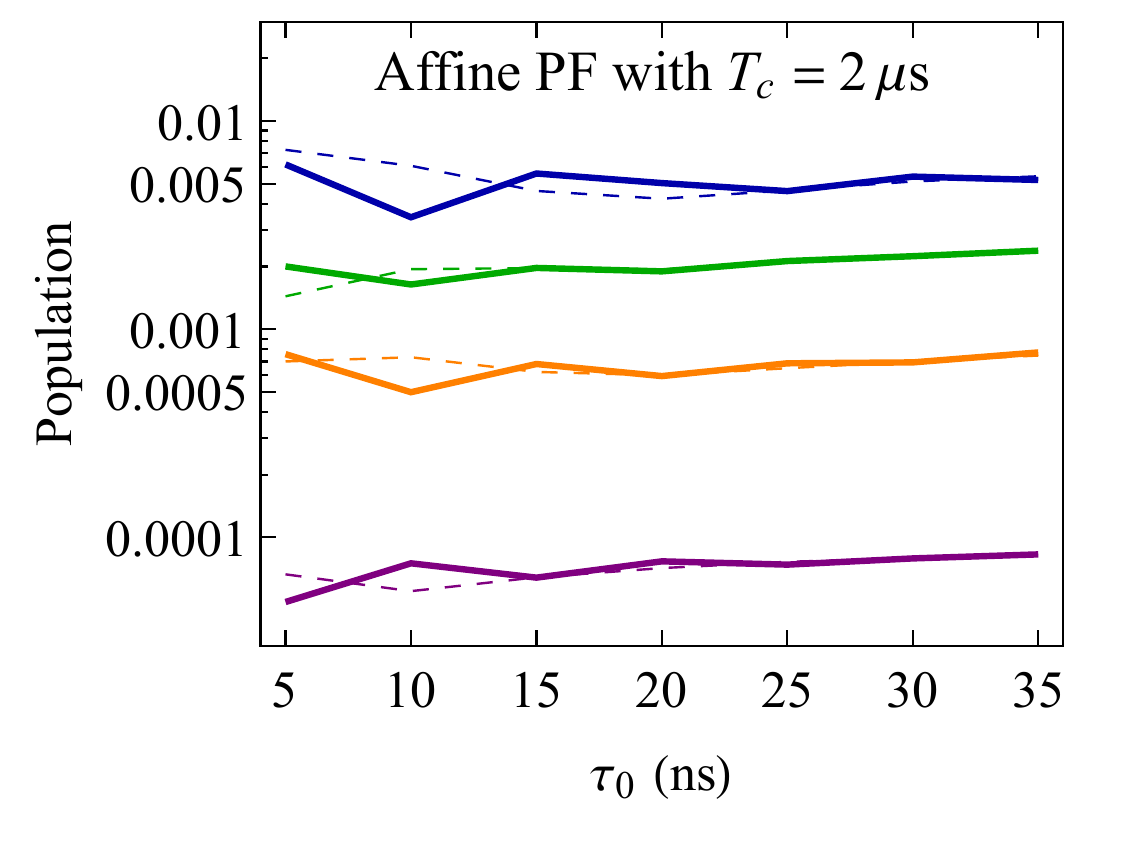}\put(-205,145){(d)}
	\vspace{-0.2in}
	\caption{ Fidelity loss of the computational states for (a) Genuine PF gate  and (b) affine PF gate due to leakage from the two pulse shapes. Coherence times of qubits are universal 200 $\mu$s while couper coherence times are much shorter with $T_1^{(c)}=T_2^{(c)}=2~\mu s$. Population of noncoumputational states $P_{ijk}(xyz)$ with initial states $|xyz\rangle$ at (c) $T_1^{(c)}=T_2^{(c)}=200~\mu s$ and (d) with $T_1^{(c)}=T_2^{(c)}=2~\mu s$. \label{fig:adia5}}
\end{figure}

In the case that coupler coherence time is much shorter than qubits e.g. $\{T_1^{(1)}, T_1^{(c)},T_1^{(2)}\}=\{T_2^{(1)}, T_2^{(c)},T_2^{(2)}\}=\{200,2,200\}$~$\mu$s, similarly we evaluate the fidelity loss of computational states in Fig.~\ref{fig:adia5}. Compared to Fig.~\ref{fig:adia}, the fidelity loss for the genuine PF gate remains almost the same while that for the affine PF gate dramatically increases. This is because if the coupler frequency is far detuning from qubits, it does not hybridize the eigenstates of qubits and thus is irrelevant to the gate performance. At the same time, it matters if the coupler frequency is close to qubits and heavily hybridizes the eigenstates. Figure~\ref{fig:adia5} (c) and (d) evaluate the population of noncomputational levels $|010\rangle,|110\rangle,|011\rangle$ initialized at $|100\rangle$/$|001\rangle$, $|101\rangle$ and $|101\rangle$, separately. One can see a significant increase in the population when the coupler coherence time is much shorter. However, using a weakly tunable coupler whose coherence time is only slightly changed and comparable with the transmon qubits does not suffer from such kind of problem, see Ref. [24].

\clearpage
\newpage
\bibliography{pfref.bib}
\end{document}